%% file: main.tex
\documentclass[twocolumn]{aastex7}
\usepackage{amsmath,amssymb}
\usepackage{mathtools}
\usepackage{color}

\newcommand*{\rev}[1]{{#1}}

\submitjournal{ApJ}
\received{October 14 2025}
\accepted{February 10 2026}

\begin{document}

\title{Conversion and Damping of Nonaxisymmetric Internal Gravity Waves in Magnetized Stellar Cores}

\author[gname='Cy',sname='David']{Cy S. David}
\affiliation{Department of Earth, Planetary, and Space Sciences, University of California, Los Angeles, CA 90095,
USA}
\email[show]{cysdavid@ucla.edu}  

\author[gname=Daniel, sname=Lecoanet]{Daniel Lecoanet}
\affiliation{CIERA, Northwestern University, Evanston, IL 60201, USA}
\email{daniel.lecoanet@northwestern.edu}

\author[gname=Pascale,sname=Garaud]{Pascale Garaud}
\affiliation{Department of Applied Mathematics, Baskin School of Engineering, University of California, Santa Cruz, CA 95064, USA}
\email{pgaraud@ucsc.edu}

\begin{abstract}
Magnetism is thought to play an important role in the evolution and dynamics of stars, though little is known about magnetic fields deep within stellar interiors. A promising avenue for probing these fields uses asteroseismic observations of global oscillations that result from the coupling of acoustic waves in the convective zone to internal gravity waves (IGWs) in the radiative interior. Recent modeling efforts implicate deep magnetic fields in the suppression of dipole mixed modes observed in 20\% of red giants and a number of high-mass main sequence stars. Previous numerical and theoretical work shows that core magnetic fields could suppress axisymmetric global modes by refracting down-going IGWs into slow-magnetosonic (SM) waves that damp at magnetic cutoff heights. Here, we extend these results to the nonaxisymmetric case, for which the IGWs and SM waves are coupled to a continuous spectrum of Alfv\'en waves (AWs). We consider a Cartesian model of the radiative interior with uniform stratification and a spatially-varying, current-free magnetic field. Using a Wentzel–Kramers–Brillouin approximation to solve for the vertical mode structure, corroborated with numerical simulations, we show that IGWs convert to up-going SM waves, which resonate with the Alfv\'en spectrum and produce mixed SM-AW modes. We find cutoff heights (as in the axisymmetric case), above which the SM/SM-AWs convert to AWs. Latitudinal variations of the background magnetic field lead to phase mixing of the AWs, resulting in rapid damping. Our results suggest that energy in both axisymmetric and nonaxisymmetric IGWs is lost via interactions with a strong magnetic field.
\end{abstract}

\section{Introduction}\label{sec:intro}
\input{1-intro}

\section{Problem setup}\label{sec:setup}
\input{2-setup}

\section{Simulation results}\label{sec:results}
\input{3-results}

\section{WKB analysis}\label{sec:wkb}
\input{4-wkb}

\section{Conclusions}\label{sec:conc}
\input{5-conc}

\begin{acknowledgments}

The authors would like to acknowledge useful conversations with Armand Leclerc, Geoff Vasil, Edgar Knobloch, Phil Marcus, Lorenzo Sironi, Matteo Cantiello, Xiaochen Sun, and Adrian Fraser regarding the slow-magnetosonic--Alfv\'{e}n waves described in this paper. CSD is supported by the NSF GRFP via award DGE-2034835. DL is partially supported by NSF AAG grant AST-2405812, Sloan Foundation grant FG-2024-21548 and Simons Foundation grant SFI-MPS-T-MPS-00007353. PG is supported by NSF AAG-2408025. This project was initiated as part of the Geophysical Fluid Dynamics summer program at the Woods Hole Oceanographic Institution, funded through NSF OCE 1829864.

\end{acknowledgments}

\section*{Code Availability}
The code used to run the simulations, perform the WKB analysis, and generate the plots in this work may be found at \rev{\url{https://github.com/cysdavid/magIGWs}}. Data from the last timestep (i.e., the equilibrated state) of the simulations are openly available at \rev{\url{https://doi.org/10.5281/zenodo.18357092}}.

\section*{Supplementary Materials}
Animated versions of Figures \ref{fig:casei}, \ref{fig:phasemixing}, and \ref{fig:caseii} can be found in the online version of this article and \rev{in a Zenodo repository (\url{https://doi.org/10.5281/zenodo.18357092})}. Additionally, a supplementary video showing the transient behavior of IVP III is included in the Zenodo repository; animations of $u_{\text{IVP}}$ and $v_{\text{IVP}}$ (decomposed by parity) are overlaid with horizontal lines (initially at the forcing height) that move vertically according to the group velocity of each WKB mode (computed using equation \ref{eqn:groupvel} in Appendix \ref{app:wkb}). The final frame corresponds to the equilibrated state in panels \textit{c} and \textit{e} of Figures \ref{fig:uwkb} and \ref{fig:vwkb}.

\begin{contribution}
CSD conducted the analytical calculations, solved the eigenvalue problems, and ran the numerical simulations. DL conceived the idea for this study and aided in both the theoretical and numerical aspects of this work. PG guided the mathematical developments, including the derivation of the reduced system and the WKB amplitude equation. All authors contributed to this manuscript.
\end{contribution}

\appendix
\input{appendix}

\bibliography{refs}{}
\bibliographystyle{aasjournalv7}

\end{document}

%% file: 1-intro.tex
Magnetic fields are thought to play an important role in stellar structure and evolution, from inflating the radii of low-mass stars \citep{Somers2020,Torres2021} to affecting chemical transport by modifying convective overshoot \citep{Brun2005}. \rev{Magnetic torques can enforce rigid-body rotation and shape the rotational evolution of massive stars \citep{spruit_birth_1998,kissin_rotation_2015,kissin_spin_2015,takahashi_modeling_2021,gouhier_angular_2022}. In high-mass main-sequence stars, magnetic fields generated in the convective core could persist into the red giant phase, when the core becomes stably-stratified \citep{braithwaite_magnetic_2017}. These ``fossil fields'' may provide the missing mechanism of angular momentum transport required to explain the modest difference in rotation rate between the contracting core and outer envelope of red giant branch (RGB) stars \citep{skoutnev_magnetic_2025}. However, while the surface magnetic fields of many stars are well-constrained by observations} \citep{Reiners2012, Wade2016}, much less is known about the magnetic fields deep within stellar interiors.

Asteroseismology \citep{Aerts2010} is one promising method to probe interior stellar magnetic fields. Weak magnetic fields produce asymmetric shifts in mode frequencies that can be calculated with perturbation theory \citep[e.g.,][]{Prat2019, Loi2020a, Dhouib2022, Mathis2023, Lignieres2024}, though \citet{Rui2024} show that perturbation theory can under-predict these shifts. Nevertheless, these types of asymmetric frequency shifts have recently been detected in tens of red-giant branch (RGB) stars \citep{li_magnetic_2022,Li2023, Deheuvels2023,Hatt2024}.

The effects of strong magnetic fields on pulsation modes is still under debate. \citet{mosser_characterization_2012,stello_prevalence_2016} found that about 20 per cent of RGB stars have lower-than-expected dipole mode amplitudes, which \citet{fuller_asteroseismology_2015} suggested was due to the presence of strong magnetic fields in their radiative cores. They argued that dipolar internal gravity waves (IGWs) would be scattered to higher wavenumbers by core magnetic fields, causing the gravity wave energy to be lost and depressing the mode amplitude. Others have used different techniques from perturbation theory to ray-tracing calculations to study the interaction of IGWs with core magnetic fields, finding that some of the wave energy is lost to magnetic waves, while some is reflected back up as IGWs, leading to a partial-suppression of the modes \citep{Loi2017, Loi2018, Loi2020c, Muller2025}. Indeed, \citet{Mosser2017} found that depressed dipolar modes still retain gravity-wave character, suggesting the IGW energy is not completely lost when they interact with magnetic fields. While no agreement has been reached regarding the efficiency of wave suppression due to strong core magnetic fields, wave suppression has been invoked to place limits on magnetic field strengths in a main-sequence star \citep{lecoanet_asteroseismic_2022} and white dwarfs \citep{Rui2025}, as well as explain tidal dissipation in planetary systems \citep{Duguid2024}.

To better understand the mechanism of IGW--magnetic field interaction, \citet{lecoanet_conversion_2017} ran two-dimensional (2D hereafter) numerical simulations in Cartesian geometry. They found that downward propagating IGWs perfectly convert to upward slow-magnetosonic (SM) waves, which develop smaller and smaller wavelengths as they propagate upward until they damp. The simulations were corroborated by a Wentzel--Kramers--Brillouin (WKB) analysis, in which the waves are assumed to be small-scale in the vertical (radial) direction, but large-scale in the horizontal (angular) direction. This contrasts with the ray-tracing analyses of \citet{Loi2020c, Muller2025} which assume small horizontal scales, which may not be appropriate for studying dipolar IGWs. This work suggested that the gravity wave energy would be completely lost when interacting with a sufficiently strong magnetic field.

In the 2D wave problem there are two different wave modes: IGWs and SM waves. However, \citet{rui_gravity_2023} pointed out that in three dimensions (3D hereafter) there exists a third wave mode, Alfv\'en waves (AWs), that may open new pathways for IGWs to interact with magnetic fields. They performed a similar WKB analysis as \citet{lecoanet_conversion_2017}, but in spherical geometry and allowing for non-axisymmetric waves. They found that axisymmetric waves have the same properties as the 2D Cartesian waves described in \citet{lecoanet_conversion_2017}. For non-axisymmetric perturbations, in addition to the IGWs and SM waves, there also exist a continuous spectrum of resonant AWs. These resonant AWs were also described in \citet{lecoanet_asteroseismic_2022}, which modeled non-axisymmetric waves in spherical geometry using the same WKB approximation for the particular case of the main-sequence B star HD 43317. However, a major limitation of both of these previous works is that it is difficult to determine how this continuous spectrum affects wave propagation using WKB theory. This leaves open the question of whether or not interactions with the continuous spectrum could allow some IGW energy to reflect toward the stellar surface.

In this work, we revisit the problem of non-axisymmetric IGWs interacting with strong magnetic fields by solving the time-dependent MHD equations, and comparing the results to a WKB analysis. Following \citet{lecoanet_conversion_2017}, we consider a simplified Cartesian model of the radiative zone of a RGB star. As in that previous work, we consider a magnetic field that varies both in $x$ (representing latitude) and $z$ (representing radius). While \citet{lecoanet_conversion_2017} assume no $y$ dependence, here we assume the waves vary as $\exp(i k_y y)$ with $k_y\neq 0$. This allows for the presence of AWs, which can couple to the IGWs and SM waves. Despite the presence of this new wave mode, we find the same physical behavior as in the  previous 2D simulations of \citet{lecoanet_conversion_2017}: there is perfect conversion from IGWs to magnetic waves (now a mix of SM waves and AWs), which damp when their wavenumbers become very large. Hence, we conclude that both axisymmetric and non-axisymmetric gravity wave energy is completely lost when interacting with a strong magnetic field.

The remainder of the paper is organized as follows. Section \ref{sec:setup} details the Cartesian model and control parameters. Simulations conducted at two values of magnetic diffusivity are presented in Section \ref{sec:results}; fine-scale features are interpreted using a simplified model of AW phase mixing. In Section \ref{sec:wkb}, we characterize the large-scale wave modes by comparing our simulation results to an asymptotic solution found using a WKB approximation in the vertical direction. Finally, Section \ref{sec:conc} summarizes our conclusions and discusses their astrophysical implications.

%% file: 2-setup.tex
\begin{figure}[t]
    \centering
    \includegraphics[width=1\linewidth]{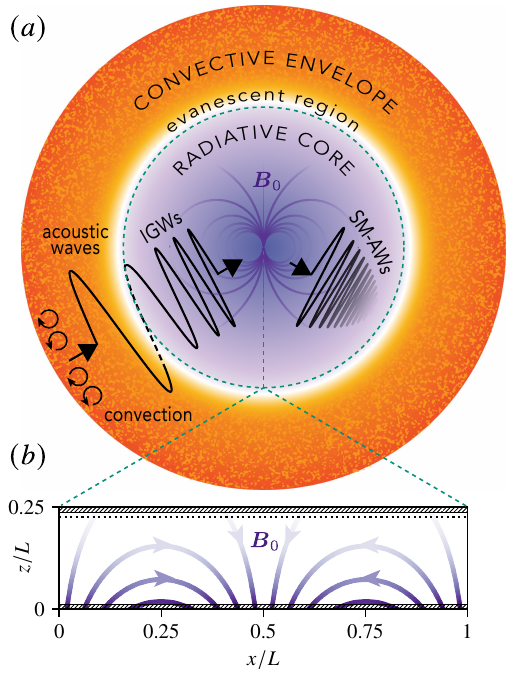}
    \caption{(\textit{a}) Schematic cross-section of the inner portion of a RGB star with convective envelope and stably-stratified radiative core. Convective turbulence in the envelope excites acoustic waves which convert to internal gravity waves (IGWs) in the radiative core. In the presence of a strong remnant core magnetic field, nonaxisymmetric IGWs convert to slow magnetosonic (SM) and resonant Alfv\'en waves (AWs). These magnetohydrodynamic waves develop fine vertical and horizontal scales as they propagate outwards, damping via diffusion. (\textit{b}) Idealized Cartesian model formed by ``unwrapping'' the radiative core, with $x$, $y$, $z$ corresponding to latitude, azimuth, and radius, respectively. The field lines for the current-free background magnetic field $\boldsymbol{B}_0$ used in this study are plotted in purple. The opacity of the plotted field lines increases with  $\lvert \boldsymbol{B}_0 \rvert$. Numerical simulations employ a sinusoidal wavemaker ($k_x = k_y = 2\pi/L$) located at the dotted line and damping layers indicated by the hatched areas.}
    \label{fig:setup}
\end{figure}

We consider a Cartesian model of the stably-stratified radiative core of a RGB star, with ``latitude'' $x \in [0,L]$, ``azimuth'' $y$, and ``radius'' $z \in [0,L_z]$ with $L_z = L/4$. Figure \ref{fig:setup}\textit{b} shows a diagram of the domain over $x$ and $z$, roughly corresponding to the dashed region of the core radiative zone of a RGB star in Figure \ref{fig:setup}\textit{a}.

The background state in our model is hydrostatic, with a stably-stratified density profile $\rho_h(z)$, such that $\partial_z \rho_h < 0$, and with a current-free magnetic field $\boldsymbol{B}_0$, such that $\boldsymbol{\nabla}\times\boldsymbol{B}_0 = \boldsymbol{0}$. Perturbations in velocity $\boldsymbol{u} = u \boldsymbol{e}_x + v \boldsymbol{e}_y + w \boldsymbol{e}_z$, pressure $p$, density $\rho$, and the magnetic field $\boldsymbol{b} = b_x \boldsymbol{e}_x + b_y \boldsymbol{e}_y + b_z \boldsymbol{e}_z$ are governed by the linearized magneto-Boussinesq equations \citep{proctor_magnetoconvection_1982}:
\begin{subequations}\label{eqn:magbouss}
    \begin{equation}\label{eqn:linmom}
         \bar{\rho} \partial_t \boldsymbol{u}+\boldsymbol{\nabla} p = -g \rho \boldsymbol{e}_z + \frac{1}{\mu_0}\left(\boldsymbol{\nabla}\times \boldsymbol{b}\right) \times \boldsymbol{B}_0,
    \end{equation}
    \begin{equation}\label{eqn:cont}
        \boldsymbol{\nabla}\boldsymbol{\cdot}\boldsymbol{u}=0,
    \end{equation}
    \begin{equation}\label{eqn:lindensity}
       \partial_t \rho = \frac{\bar{\rho} N^2}{g}w,
    \end{equation}
    \begin{equation}\label{eqn:lininduction}
        \partial_t \boldsymbol{b} = \boldsymbol{\nabla}\times(\boldsymbol{u}\times\boldsymbol{B}_0) + \eta \nabla^2 \boldsymbol{b},
    \end{equation}
    \begin{equation}\label{eqn:gauss}
        \boldsymbol{\nabla}\boldsymbol{\cdot}\boldsymbol{b}=0,
    \end{equation}
\end{subequations}
where $\bar{\rho}$ is the mean density (a constant), $g$ is the gravitational acceleration, $\mu_0$ is the magnetic permeability of free space, and $\eta$ is the magnetic diffusivity. The dimensionless form of these equations used to derive our theoretical results are provided in Appendix \ref{app:ndim}.

We assume a constant Brunt-V\"ais\"al\"a frequency $N = \sqrt{-g \bar{\rho}^{-1} \partial_z \rho_h}$. For simplicity, we ignore rotation, following \cite{lecoanet_conversion_2017} and \cite{rui_gravity_2023}. This choice may be somewhat justified in the case of RGB stars, since the frequency of observed dipole modes $\omega$ is much faster than the rotational frequency $\Omega$ ($\omega/\Omega \sim 20$, \citealt{stello_prevalence_2016,gehan_core_2018}). In our simulations, magnetic diffusion is retained to regularize sharp features in the simulations associated with AWs (detailed in the next section), while viscous and radiative diffusion are neglected. Though this choice does not reflect the actual ordering of diffusivities in stars (typically, the radiative diffusivity is many orders of magnitude larger than $\eta$), it is a numerical convenience that allows the AWs to be resolved while minimally affecting the propagation of IGWs.

\begin{deluxetable*}{rcccccccccc}
\tablewidth{0pt}
\tablecaption{Simulation parameters\label{tab:simparams}}
\tablehead{
\colhead{Simulation} & \colhead{($N_x$, $N_z$)} & \colhead{$\textit{Fr}$} & \colhead{$\Gamma$} & \colhead{$k_x L = k_y L$} & \colhead{$\textit{Lu}$} & \colhead{$\textit{Lu}_x$} & \colhead{$z_0/L$} & \colhead{$\Delta z/L$} & \colhead{$s/L$} & \colhead{$\Delta t/\omega^{-1}$}
}
\startdata
IVP I     & (2048, 1536)   & 0.025          & 0.1      & $2 \pi$         & $6.25 \times 10^4$ & —               & 0.225   & 0.00416667   & 0.0125 & 0.0111     \\
IVP II    & (4096, 3072)   & 0.025          & 0.1      & $2 \pi$         & $6.25 \times 10^5$ & —               & 0.225   & 0.00416667   & 0.0125 &  0.00278          \\
IVP III   & (2048, 1536)   & 0.025          & 0.1      & $2 \pi$         & $6.25 \times 10^4$ & $12.5$ & 0.225   & 0.00416667   & 0.0125 & 0.0111     \\ 
\enddata
\tablecomments{Our simulations have a domain size of $L_x = L$ by $L_z = L/4$. We use $N_x$ ($N_z$) modes in the horizontal (vertical) directions and a uniform timestep $\Delta t$. Waves are forced at $z=z_0$ over a region with thickness $\Delta z$, and they are damped within layers of size $s$. The forcing frequency and horizontal wavenumbers are denoted by $\omega$, $k_x$, and $k_y$, respectively. The Froude number $\textit{Fr}$, magneto-gravity ratio  $\Gamma$, and Lundquist number $\textit{Lu}$ are defined in Section \ref{sec:setup}. As discussed in Section \ref{sec:wkb}, we use anisotropic diffusion in IVP III, where the strength of diffusion in the $x$ direction is measured with $\textit{Lu}_x = \omega l_z^2/\eta_x$.}
\end{deluxetable*}

An important control parameter for the waves is the  Froude number, defined as
\begin{equation}
    \textit{Fr} = \frac{\text{wave inertia}}{\text{buoyancy}} =\frac{\omega}{N}.
\end{equation}
In the absence of magnetic fields, the Froude number controls the vertical scale of oscillations, as can be seen by rearranging the dispersion relation for pure IGWs:
\begin{equation}\label{eqn:igwdisprel}
    k_z^2 = (k_x^2 + k_y^2)(\textit{Fr}^{-2} - 1),
\end{equation}
where $k_z$ is the ``radial'' wavenumber. In the limit of strong stratification, which we define as $\textit{Fr} \ll 1$, the vertical wavelength scales as $2\pi/k_z \sim l_z \equiv \textit{Fr} L$. This anisotropy is expected to hold for the suppressed dipole modes in RGB stars ($\textit{Fr} \sim 10^{-1}$, \citealt{stello_prevalence_2016,montalban_testing_2013}) and is consistent with the Boussinesq approximation, for which vertical scales must be much smaller than the pressure scale height (the latter being comparable to $L$).

\rev{We impose a background magnetic field that varies slowly in $x$ and has} vanishing vertical component at the ``equator'' ($x/L = 0.25, 0.75$):
\begin{equation}\label{eqn:backgroundfield}
    \boldsymbol{B}_0 = \mathcal{B} e^{-2\pi z/L}\left[\sin(2\pi x/L)\boldsymbol{e}_x + \cos(2\pi x/L)\boldsymbol{e}_z\right].
\end{equation}
Field lines associated with $\boldsymbol{B}_0$ are plotted in purple in Figure \ref{fig:setup}\textit{b}. The assumed exponential decay with $z$ ensures that the field is current-free (i.e., $\boldsymbol{J}_0 = \mu_0^{-1}\boldsymbol{\nabla} \times \boldsymbol{B}_0 = \boldsymbol{0}$) \rev{in the simulated domain (with the exception of damping layers introduced later in this section) and is} thus stable to perturbations. Hence, $\boldsymbol{B}_0$ may be thought of as the Cartesian equivalent of a dipole field. 

\rev{Though fossil magnetic fields in RGB stars likely have a toroidal component ($B_\phi$) of comparable (if not greater) magnitude than their poloidal components ($B_r$, $B_\theta$) \citep{braithwaite_stable_2006,braithwaite_nonaxisymmetric_2008, braithwaite_axisymmetric_2009,duez_relaxed_2010}, comparison of the terms in $\boldsymbol{B} \cdot \boldsymbol{\nabla} \boldsymbol{b}$ reveals that $B_{\phi}$ has negligible effect on the IGWs considered here so long as}
\begin{equation}\label{eqn:toroidalcondition}
    \lvert B_\phi \rvert \ll \left\lvert \frac{r k_r B_r \sin \theta}{m}\right\rvert,
\end{equation}
\rev{where $r$ is radius, $\theta$ is colatitude, $k_r$ is the radial wavenumber, and $\lvert m \rvert = 1$ for dipolar waves. If $
\lvert B_\phi \rvert \sim \lvert B_r \sin \theta \rvert$, condition (\ref{eqn:toroidalcondition}) is easily satisfied since we are restricting our analysis to waves where $k_r r \sim \textit{Fr}^{-1} \gg 1$ away from the center of the core. While dipolar IGWs with short radial wavelengths interact most strongly with the radial background field component if all components are of similar magnitude, (\ref{eqn:toroidalcondition}) shows that large toroidal fields may still enter the problem at leading order. Nevertheless, in this investigation, we set the toroidal component of our model's background field to zero (i.e., $\boldsymbol{B}_0 \cdot \boldsymbol{e}_y =0$) for simplicity. Finally, though the ``latitudinal'' component $\boldsymbol{B}_0 \cdot \boldsymbol{e}_x$ has only a small influence on the waves of interest, it is retained to permit gradients in $\boldsymbol{B}_0$ without introducing monopoles or currents.}

The presence of a magnetic field influences wave motion via the Lorentz force and Ohmic diffusion. The latter is characterized by the local-scale Lundquist number,
\begin{equation}
    \textit{Lu} = \frac{\text{wave inertia}}{\text{magnetic diffusion}} = \frac{\omega l_z^2 }{\eta} = \frac{\omega^3 L^2}{N^2 \eta},
\end{equation}
and is expected to have negligible effect on IGWs in the RGB core ($\textit{Lu} \sim 10^{16}\text{–}10^{20}$, \citealt{stello_prevalence_2016,montalban_testing_2013,griffiths_magneto-rotational_2022}).

The strength of the magnetic field is measured in terms of
the dimensionless ``magneto-gravity ratio''
\begin{equation}
    \Gamma = \frac{\text{Lorentz force}}{\text{wave inertia}} = \left(\frac{\omega_{\textit{MG}}}{\omega}\right)^2,
\end{equation}
where
\begin{equation}
    \omega_{\textit{MG}} = \left( \frac{\mathcal{B} N}{L\sqrt{\mu_0 \bar{\rho}}}\right)^{1/2}
\end{equation}
is the magneto-gravity frequency \citep{fuller_asteroseismology_2015}. For the stellar cores of interest, we cannot determine the value of $\Gamma$ \textit{a priori} since $\mathcal{B}$ is not known at the outset. Instead, we follow \cite{fuller_asteroseismology_2015} in assuming that the background magnetic field is strong enough to influence IGWs but not so strong as to inhibit horizontal ($x$,$y$) perturbations, $\boldsymbol{u}_H$, $\boldsymbol{b}_H$. Accordingly, we assume that
\begin{equation}
    \lvert \partial_t \boldsymbol{u}_H \rvert \gtrsim \left\lvert \frac{\boldsymbol{B}_0 \boldsymbol{\cdot} \boldsymbol{e}_z}{\mu_0 \bar{\rho}} \partial_z \boldsymbol{b}_H\right\rvert.
\end{equation}
Using $l_z = \omega L/N$ as the scale for the vertical gradient, $\mathcal{U}$ as the (arbitrary) scale of $\boldsymbol{u}_H$, and $\mathcal{U} \sqrt{\mu_0 \bar{\rho}}$ as the scale of $\boldsymbol{b}_H$, this implies that
\begin{equation}
    \omega \mathcal{U} \gtrsim \frac{\mathcal{B}}{\mu_0 \bar{\rho}}  \frac{1}{l_z} \mathcal{U} \sqrt{\mu_0 \bar{\rho}},
\end{equation}
and therefore
\begin{equation}
    \Gamma = \frac{\mathcal{B} N}{\omega^2 L \sqrt{\mu_0 \bar{\rho}}} \lesssim 1.
\end{equation}

The magneto-gravity ratio $\Gamma$ used in this work is similar to the ``depth parameter'' $a$ in \cite{rui_gravity_2023}. However, $a$ is proportional to the local magnetic field strength (and may be used as a vertical coordinate) whereas $\Gamma$ is proportional to the strength $\mathcal{B}$ of the background field at the base of the domain (and is thus a constant). In our problem, this is enough to characterize $\lvert \boldsymbol{B}_0 \rvert$ everywhere since $\lvert \boldsymbol{B}_0 \rvert = \mathcal{B} \exp(-2\pi z/L)$ .

To study the interaction of IGWs with magnetic fields, we numerically solve initial value problems (IVPs) at $\textit{Fr} = 0.025$ and $\Gamma = 0.1$, in which waves are forced at a fixed frequency $\omega$ and horizontal wavenumbers $k_x = k_y = 2 \pi/L$. Though we cannot reach astrophysical values of $\textit{Lu}$, we show that the relevant physics already occur at more moderate values of $\textit{Lu}$ by comparing two cases with $\textit{Lu} = 6.25 \times 10^4$ (IVP I) and $\textit{Lu} = 6.25 \times 10^5$ (IVP II).

We assume that all dependent quantities can be expressed as $q(x,y,z,t) = \Re\{q'(x,z,t) \exp({i k_y y})\}$ (so $\partial_y \equiv i k_y$), and we solve the remaining partial differential equations in $x,z,t$ using the Dedalus pseudospectral code \citep{burns_dedalus_2020} with ($N_x$, $N_z$) complex Fourier modes in both directions. A two stage, second-order implicit–explicit Runge–Kutta scheme \citep{ASCHER1997151} with a uniform timestep $\Delta t$ is used to evolve the flow from an initially quiescent state. The timestep, resolution, and other simulation parameters are reported in Table \ref{tab:simparams}.

To excite ``dipolar'' IGWs with fixed frequency $\omega$ and latitudinal wavenumber $k_x = 2 \pi/L$, we add a forcing term \rev{proportional to}
\begin{equation}\label{eqn:forcing}
    F = \exp\left[-\frac{(z-z_0)^2}{\Delta z^2}\right]\exp\left[i(k_x x + k_y y - \omega t)\right],
\end{equation}
to the right-hand side of the density equation (\ref{eqn:lindensity}), where $\Delta z = L/240$ is the vertical width of the forcing and $z_0 = 9 L/40$ is the forcing height (dotted line in Figure \ref{fig:setup}\textit{b}). We set the azimuthal wavenumber of the forcing to $k_y=k_x=2\pi/L$ by analogy with nonaxisymmetric dipole modes, which have $\lvert m \rvert = \ell = 1$.

In order to avoid unphysical reflections of the waves at the top and bottom boundaries of the domain, we add damping layers near $z = 0$ and $z = L_z$ (hatched regions in Figure \ref{fig:setup}\textit{b}). These layers are implemented by adding linear damping terms $-\bar{\rho}\omega D \boldsymbol{u}$, $-\omega D \rho$, and $-\omega D \boldsymbol{b}$ to the right hand sides of (\ref{eqn:linmom}), (\ref{eqn:lindensity}), and (\ref{eqn:lininduction}), respectively. The \rev{masking function}
\begin{equation}\label{eqn:damping}
    D(z) = \frac{1}{2}\left[ \tanh\left(\frac{z - L_z + s}{\Delta z}\right) + \tanh\left(\frac{s - z}{\Delta z}\right) + 2\right]
\end{equation}
ensures that perturbations are damped on timescales longer than $\omega^{-1}$ \rev{within layers of size }$s = L/80$.

The use of Fourier modes in $z$ requires the background magnetic field to be periodic. We achieve this by multiplying $\boldsymbol{B}_0$ by $-\frac{1}{2}[ \tanh\left(({z - L_z + s_B})/{\Delta z_B}\right)$ $+ \tanh\left(({s_B - z})/{\Delta z_B}\right)]$, where $s_B = s/3$ and $\Delta z_B = \Delta z/4$ so that the field is tapered to $\boldsymbol{0}$ within the damping layers at the top and bottom of the domain. \rev{After applying this masking function, $\boldsymbol{B}_0$ is current-free only within the undamped regions of the domain.}

The simulations are integrated until $t = 1000 \omega^{-1}$, long after the time at which the mean wave energy per unit volume
\begin{equation}\label{eqn:meanenergy}
E  = \frac{k_y}{2\pi L_x L_z}\int_0^{L_z}\int_{-\pi/{k_y}}^{\pi/{k_y}}\int_0^{L_x} \mathcal{E}\mathrm{d}x\mathrm{d}y\mathrm{d}z,
\end{equation}
saturates ($E$ reaches 99 per cent of its final value at least 75 wave periods before the end of each simulation). In (\ref{eqn:meanenergy}), the local wave energy density is
\begin{equation}
    \mathcal{E} = \frac{1}{2}\left(\bar{\rho}{ \boldsymbol{u}\cdot\boldsymbol{u}} + \frac{g^2 \rho^2}{\bar{\rho} N^2}+\frac{\boldsymbol{b} \cdot \boldsymbol{b}}{\mu_0} \right).
\end{equation}

%% file: 3-results.tex
\begin{figure*}[t]
    \centering
    \includegraphics[width=\linewidth]{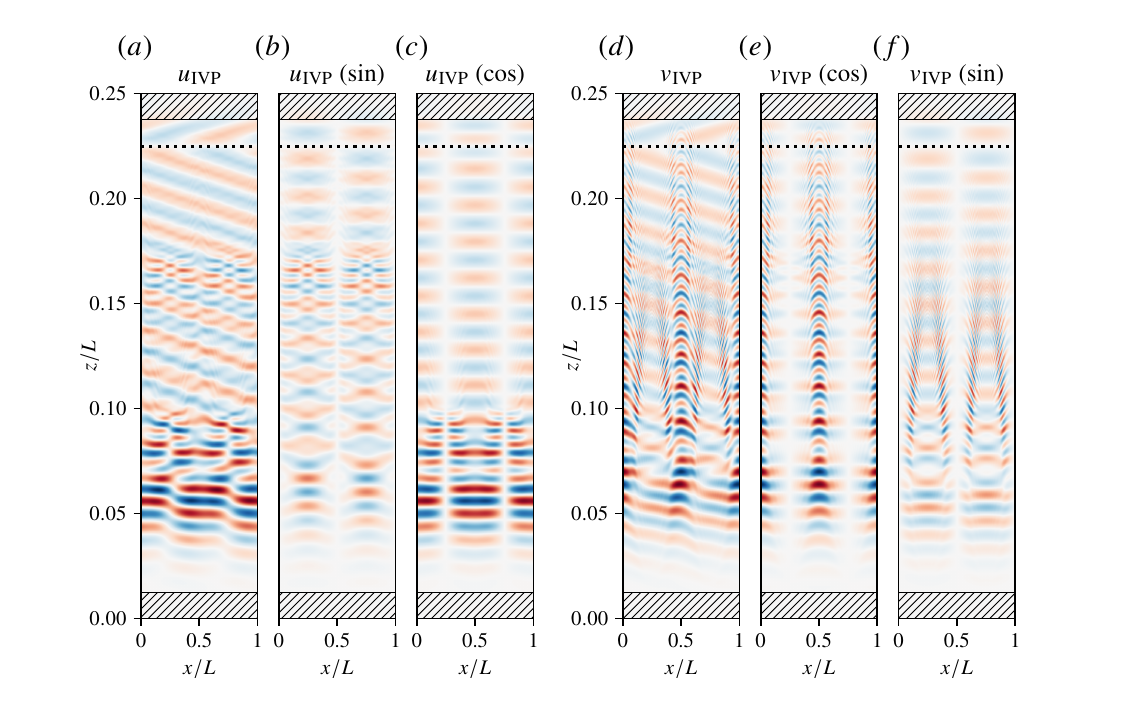}
    \caption{Snapshots of the latitudinal ($x$) and azimuthal ($y$) velocity perturbations ($u_{\text{IVP}}$, $v_{\text{IVP}}$) from IVP I ($\textit{Lu} = 6.25 \times 10^4$, $\textit{Fr} = 0.025$, $\Gamma =0.1$) after the simulation has equilibrated. (\textit{a}) Plot of $u_{\text{IVP}}$ over ($x$,$z$) plane located at $y = 0$. IGWs are forced at the dotted line and damp in the hatched regions. (\textit{b})–(\textit{c}) Sine and cosine-parity components of $u_{\text{IVP}}$, respectively. (\textit{d}) The azimuthal velocity component, $v_{\text{IVP}}$.  (\textit{e})–(\textit{f}) Cosine and sine-parity components of $v_{\text{IVP}}$, respectively. A \rev{42-second }animated version of this figure (available in the online article and at \rev{\url{https://doi.org/10.5281/zenodo.18357092}}) shows the transient behavior of IVP I \rev{over $1000/(2\pi)$ wave periods.} The fine shingle-like features (associated with AWs) in panel \textit{e} (above $z/L \approx 0.17$) and panel \textit{f} (above $z/L \approx 0.1$)\rev{, emerge at 00:10 and 00:08 (respectively) and} propagate upwards. \rev{The simulation equilibrates by 00:14}.}
    
    \label{fig:casei}
\end{figure*}
\begin{figure}
    \centering
    \includegraphics[width=\linewidth]{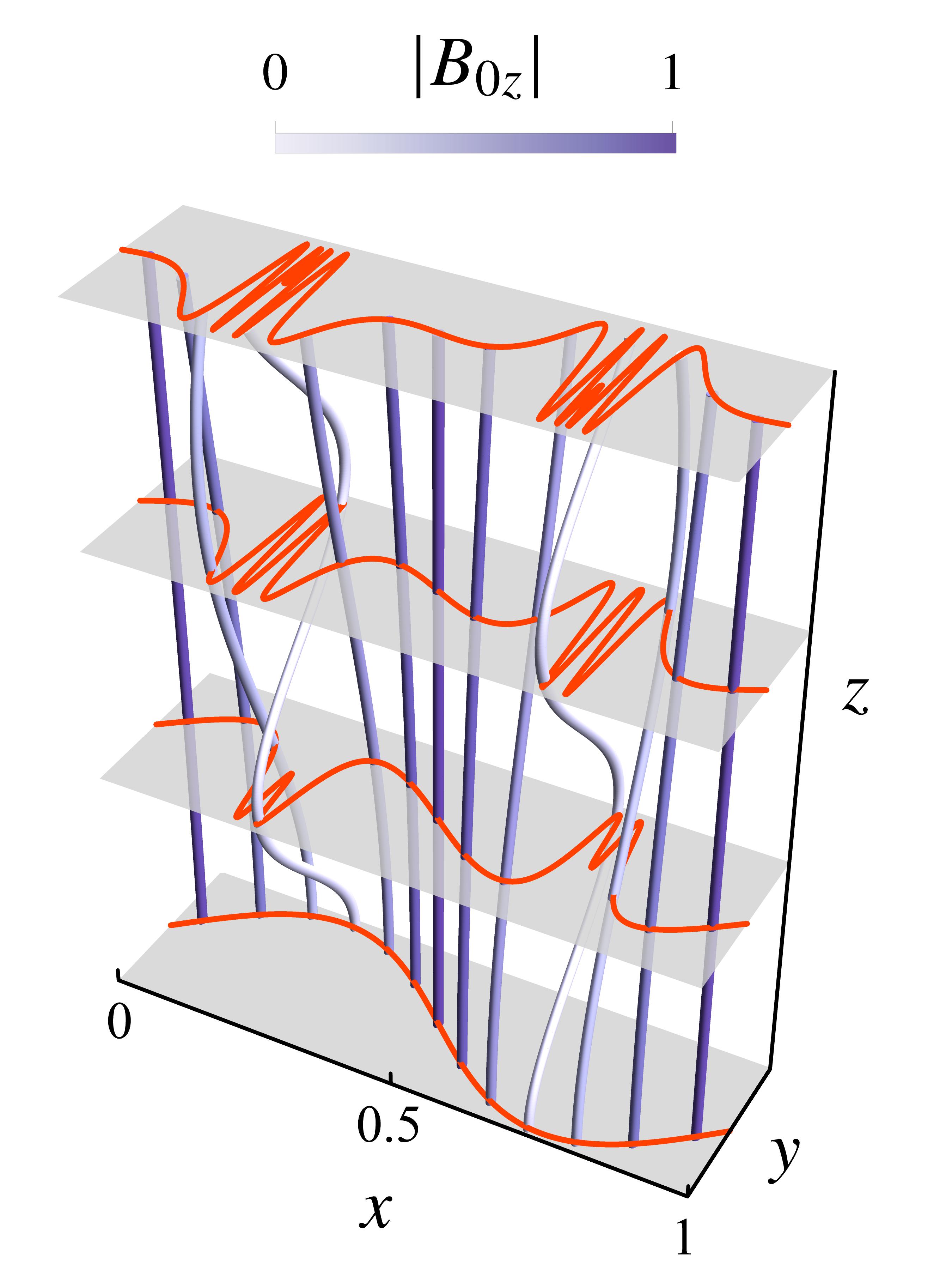}
    \caption{Three-dimensional rendering of magnetic field lines $B_{0 z}\boldsymbol{e}_z + b_y\boldsymbol{e}_y$ perturbed by AWs in an axisymmetric ($k_y = 0$) toy problem where the background magnetic field strength varies according to $B_{0 z}^2 = \cos^2(2\pi x) + 0.01$; darker shades of purple correspond to higher $\lvert B_{0 z} \rvert$. The open field lines are sinusoidally forced in the azimuthal ($y$) direction at their base with fixed frequency $\omega$. Each field line oscillates with a different vertical wavenumber given by (\ref{eqn:toyalfvenwavenumber}), leading to phase mixing that increases with height ($z$). Profiles of the azimuthal fluid displacement $\xi_y$ as a function of $x$ are overlaid in orange. A \rev{26-second }animated version of this figure (available in the online article and at \rev{\url{https://doi.org/10.5281/zenodo.18357092}}) shows the oscillation of the magnetic field lines according to (\ref{eqn:toydisplacement}).}
    \label{fig:phasemixing}
\end{figure}
\begin{figure*}[t]
    \centering
    \includegraphics[width=1\linewidth]{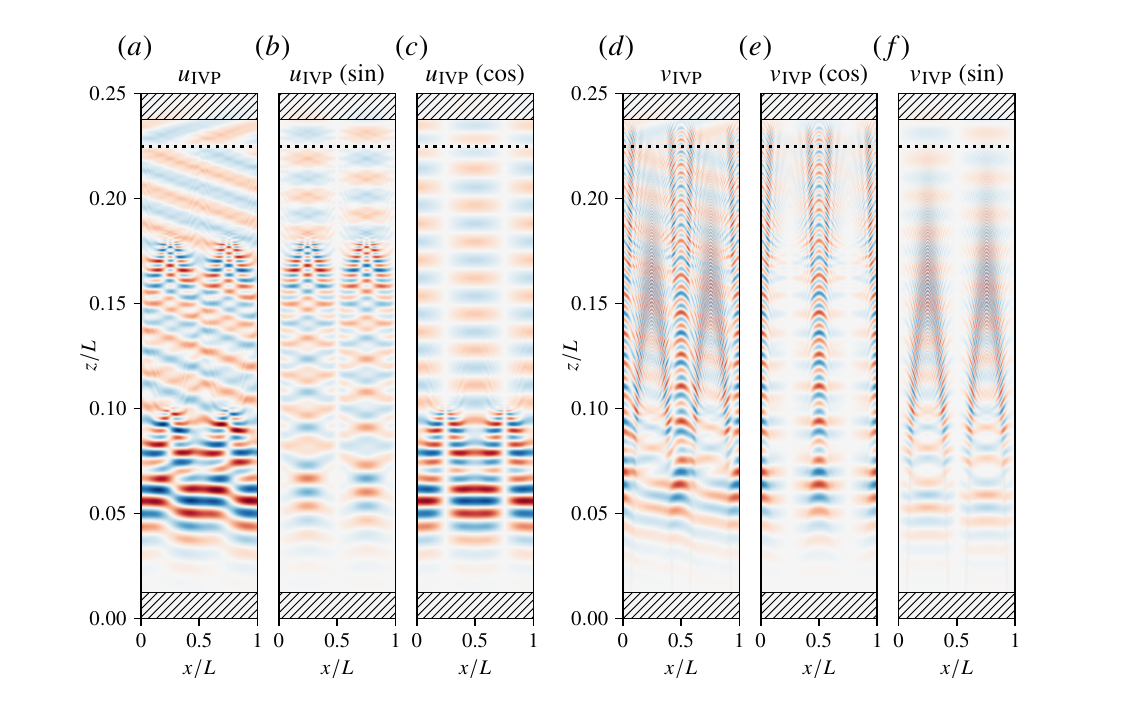}
    \caption{Snapshots of the latitudinal ($x$) and azimuthal ($y$) velocity perturbations ($u_{\text{IVP}}$, $v_{\text{IVP}}$) from IVP II ($\textit{Lu} = 6.25 \times 10^5$, $\textit{Fr} = 0.025$, $\Gamma =0.1$) after the simulation has equilibrated. Panel descriptions are the same as in Figure \ref{fig:casei}. A \rev{42-second }animated version of this figure (available in the online article and at \rev{\url{https://doi.org/10.5281/zenodo.18357092}}) shows the transient behavior of IVP II \rev{over $1000/(2\pi)$ wave periods. As in Figure \ref{fig:casei}, the AWs in panels $\textit{e}$ and $\textit{f}$ emerge at 00:10 and 00:08, respectively. The simulation equilibrates by 00:22.}}
    \label{fig:caseii}
\end{figure*}

Figure \ref{fig:casei} shows snapshots of IVP I ($\textit{Lu} = 6.25 \times 10^4$) in the equilibrated state. Panel \textit{a} presents the latitudinal flow $u$ in the $(x,z)$ plane located at $y=0$; the corresponding azimuthal flow $v$ is shown in panel \textit{d}. The domain is vertically stretched to better show the rapid oscillations in $z$, as anticipated from dimensional analysis ($l_z = \textit{Fr} L$, $\textit{Fr} = 0.025 \ll 1$). Hatched regions indicate the damping layers.

Since the $x$ and $z$ components of the background magnetic field $\boldsymbol{B}_0$ have distinct parities, the sine- and cosine-parity components of $u$ and $v$ behave differently upon interacting with $\boldsymbol{B}_0$; it is therefore instructive to analyze these components separately. To satisfy incompressibility, $u$ and $v$ must have opposite parity in $x$: the sine component of $u$ (shown in panel \textit{b}) is associated with the cosine component of $v$ (shown in panel  \textit{e}), and the cosine component of $u$ (panel \textit{c}) is associated with the sine component of $v$ (panel \textit{f}). Hereafter, we shall refer to waves associated with sine-parity (cosine-parity) latitudinal flow $u$ as sine-parity (cosine-parity) waves.

The forcing (dotted line at $z_0/L = 0.225$) excites IGWs with both sine and cosine parity (checkerboard features in Figure \ref{fig:casei}\textit{b},\textit{c},\textit{e},\textit{f}). At early times in the simulation, these IGWs travel downwards into a region of increasingly strong magnetic field. Below $z/L \approx 0.05$ (still many wavelengths above the bottom damping layer), the field becomes too strong to permit wave propagation, and now causes the down-going waves to become evanescent instead (see Section \ref{sec:wkb} for more detail). Above $z/L \approx 0.05$, the strong magnetic field causes the waves to reflect and propagate back in the upwards direction. These up-going waves (of both sine and cosine parities) have different horizontal structures than the IGWs forced at the top of the domain.

The latitudinal flow $u$ associated with the sine-parity up-going wave (Figure \ref{fig:casei}\textit{b}) develops shorter and shorter vertical scales as the magnetic field weakens with height, until it vanishes near $z/L \approx 0.17$. This apparent cut-off height resembles the SM cutoff heights in the axisymmetric case \citep{lecoanet_conversion_2017}, as discussed further in Section \ref{sec:wkb}. However, for $k_y \neq 0$, the coupling of the latitudinal flow $u$ to the azimuthal flow $v$ introduces additional complexity. The sine-parity up-going wave is associated with azimuthal flow $v$ (shown in panel \textit{e}) that exhibits large-scale ridges centered at $x/L = 0, 0.5, 1$. The amplitude of these ridges decays gradually above the cutoff height near $z/L \approx 0.17$ (in contrast to the concurrent sharp decay in $u$). At the cutoff height, energy in the up-going wave is transferred from $u$ to $v$, coinciding with the growth of fine-scale tilted features in $v$ on either side of each ridge.

The latitudinal flow $u$ associated with the cosine-parity up-going wave (Figure \ref{fig:casei}\textit{c}) also vanishes at a cutoff height located this time around $z/L \approx 0.1$. At this point, energy is transferred to the sine-parity azimuthal flow (shown in panel \textit{f}), which develops features with extremely fine horizontal scales.

The axisymmetric simulations of \cite{lecoanet_conversion_2017} also contain small-scale oscillations, for which there was no physical interpretation in that work. Those features are extremely limited in latitudinal extent (concentrated around $x/L = 0.25, 0.75$ where the vertical component of the background field vanishes) in contrast with the fine-scale features in Figure \ref{fig:casei}\textit{e},\textit{f}. As discussed further below, we interpret these features (in the present simulations) as AWs.

AWs are known to develop small horizontal scales as a consequence of phase mixing (due to the variation of Alfv\'en speed with $x$) and rapidly damp via Ohmic diffusion \citep[e.g.,][]{hasegawa_plasma_1974,heyvaerts_coronal_1983}. To show how phase mixing occurs in our simulations, we present a `toy' model (after \citealt{heyvaerts_coronal_1983}), involving a purely vertical background magnetic field $B_{0 z}(x)\boldsymbol{e}_z$ that varies with $x$. We consider open field lines and an axisymmetric forcing that displaces fluid at the base of the domain $z=z_b$ by $\xi_b(x)\exp(-i\omega t)\boldsymbol{e}_y$, where $\xi_b(x)$ is an arbitrary displacement profile.

In the absence of diffusion, each magnetic field line can move independently in the $y$ direction and thus supports AWs associated with azimuthal fluid displacement
\begin{equation}\label{eqn:toydisplacement}
    \xi_y = \frac{v}{-i\omega} = \frac{b_y}{i k_A B_{0 z}} = \xi_b(x)\exp[i k_A(x) (z-z_b)-i\omega t]
\end{equation}
where
\begin{equation}\label{eqn:toyalfvenwavenumber}
    k_A(x) = \frac{\omega \sqrt{\mu_0 \bar{\rho}}}{ \lvert B_{0 z}(x) \rvert}
\end{equation}
is the local Alfv\'en wavenumber.

To illustrate the resultant phase mixing process, we consider a basal displacement profile $\xi_b(x) = \sin(2\pi x)$ and background magnetic field with $B_{0 z}^2 = \cos^2(2\pi x) + 0.01$ to ensure that $k_A(x)$ remains bounded for all $x$. Figure \ref{fig:phasemixing} shows magnetic field lines $B_{0 z}\boldsymbol{e}_z + b_y\boldsymbol{e}_y$ (purple curves) associated with the forced AWs. Profiles of the azimuthal displacement $\xi_y$ over $x$ are overlaid in orange at four different heights. Since stiffer field lines (higher $\lvert B_{0 z}(x) \rvert$, darker purple curves) oscillate with longer wavelengths (lower $k_A$), field lines that are in phase at one height become out of phase higher up. This process of phase mixing produces small horizontal scales in $\xi_y$ (and thus $v$) that become finer with height (compare the top and bottom orange curves). The finest scales are concentrated where $k_A$ changes most rapidly with $x$, around minima in $\lvert B_{0 z}(x) \rvert$ (at $x = 0.25, 0.75$).

Phase mixing in IVP I produces similar features in the azimuthal flow. For the cosine-parity wave, Figure \ref{fig:casei}\textit{f} shows fine scales in $v$ above $z/L \approx 0.15$, concentrated around $x/L = 0.25, 0.75$ (near which $B_{0 z}$ changes rapidly), as in the toy problem. However, our simulations contain further complexity: the background field is not purely vertical, magnetic diffusion is present, and the waves are nonaxisymmetric such that azimuthal and latitudinal motions are coupled via continuity. Further, sharp features in $v$ (not due to phase mixing) are already present in the up-going large-scale wave modes below their cutoff heights.

For the cosine-parity wave, these sharp features (the tilted ``shingles’’ in Figure \ref{fig:casei}\textit{f} below $z/L \approx 0.1$ and away from $x/L = 0.25, 0.75$) are located at critical layers where the large-scale wave mode resonates with the Alfv\'en spectrum, as will be demonstrated in Section \ref{sec:wkb}. At the cutoff height (near $z/L \approx 0.1$), the large-scale wave mode acts analogously to the basal forcing in the toy problem, transferring energy to the phase-mixing AWs. In the toy problem, the sinusoidal structure of the forcing is imparted to the AWs. Similarly, in the simulation, the AWs just above $z/L \approx 0.1$ retain the critical layer shingles present in the large-scale wave mode.

The large-scale up-going wave with sine parity also converts to AWs above its cutoff height (near $z/L \approx 0.17$). As in the cosine-parity case, features of the large-scale wave (i.e., the ridges centered at $x/L = 0,0.5,1$ in Figure \ref{fig:casei}\textit{e}) persist in the azimuthal flow associated with the AWs and gradually decay with height. The effect of phase mixing is weaker than before since $v$ vanishes at  $x/L = 0.25, 0.75$ for the sine-parity waves. Nonetheless, thin tilted shingles associated with phase mixing are still visible next to the ridge features in Figure \ref{fig:casei}\textit{e}, and they grow finer towards $x/L = 0.25, 0.75$ as expected.

Fine-scale AWs of both parities are efficiently damped by magnetic diffusion. In order to isolate the effects of resistivity on the AWs and large-scale modes of oscillation, we run a second IVP (IVP II) and decrease the magnetic diffusivity so that $\textit{Lu} = 6.25 \times 10^5$.

Figure \ref{fig:caseii} shows $u$ and $v$ decomposed by parity for IVP II. We see that the large scale features observed in IVP I are still present in IVP II, and are mostly unaffected by the decrease in magnetic diffusivity. Similarly, AWs emerge as before at $z/L \approx 0.17$ and $z/L \approx 0.1$ for the sine and cosine parities (Figure \ref{fig:caseii}\textit{e} and \textit{f}), respectively. However, $v$ develops significantly finer scales in IVP II than in IVP I, since diffusion sets the minimum length-scale of AWs \citep[e.g.,][]{poedts_efficiency_1990}.

\rev{The scaling behavior of this resistive length with the magnetic diffusivity may be found by modifying the toy problem above to include ohmic diffusion. As before, a periodic axisymmetric forcing at the base of the domain $z=z_b$ drives azimuthal flow with $v(x,z_b,t) = v_b(x)\exp(-i\omega t)$. But, for better comparison to our simulations, we now consider an exponentially-decaying ``poloidal'' background field $\boldsymbol{B}_0 = B_{0 x}\boldsymbol{e}_x+B_{0 z}\boldsymbol{e}_z$ of the form}
\begin{subequations}
\begin{equation}
    B_{0 x} = f(x)\exp(-2\pi z/L),
\end{equation}
   \begin{equation}
    B_{0 z} = h(x)\exp(-2\pi z/L).
\end{equation} 
\end{subequations}

\rev{For simplicity, we assume that the horizontal component $B_{0 x}$ is small compared to the vertical component $B_{0 z}$ (and is thus rapidly-varying in $x$ such that $\boldsymbol{\nabla} \cdot \boldsymbol{B}_0 = 0$). Then, axisymmetric, azimuthal motions propagate nearly vertically as AWs, governed by the $y$-components of (\ref{eqn:linmom}) and (\ref{eqn:lininduction}). Following the WKB approach of \cite{de_moortel_phase_2000}, we approximate the azimuthal velocity and magnetic field perturbations as}
\begin{subequations}
\begin{equation}\label{eqn:vansatz}
    v(x,z,t) \sim v_0(x,z)\exp\left[i \theta_A(x,z)  - i \omega t\right],
\end{equation}
\begin{equation}\label{eqn:byansatz}
    b_y(x,z,t)   \sim b_{y 0} (x,z)\exp\left[i \theta_A(x,z)  - i \omega t\right],
\end{equation}
\end{subequations}
\rev{where $\theta_A(x,z_b) = 0$ and $v_0(x,z_b) =  v_b(x)$.}

\rev{The vertical wavenumber $k_A = \partial_z \theta_A$ is assumed to be large ($k_A L \sim \varepsilon^{-1}$, $\varepsilon \ll 1$) and to vary slowly with both $x$ and $z$, since the vertical component of the background field varies over the system scale $L$ in both directions. The amplitude functions $v_0$ and $b_{y 0}$ also vary over the system scale, with the $x$-dependence controlled by the basal displacement profile and the $z$-dependence determined by the rate of damping. Further, we assume that the diffusion acting over the vertical scales of the AWs is small ($\eta k_A^2/\omega \sim \varepsilon$) such that the dispersion relation is ideal (to leading order) and resistive effects only enter the problem through the amplitude equation (at the next order in $\varepsilon$). This is equivalent to requiring the system-scale Lundquist number to follow $\textit{Lu}_L = \omega L^2/\eta \sim \varepsilon^{-3}$. Finally, we restrict our analysis to latitudes where the horizontal magnetic field is sufficiently small: $B_{0 x}/B_{0 z} \sim \varepsilon^2$.}

\rev{Then, at the leading order in $\varepsilon$ we find that}
\begin{subequations}
    \begin{equation}\label{eqn:kA}
        k_A(x,z) = \frac{\omega \sqrt{\mu_0 \bar{\rho}}}{ \lvert B_{0 z}(x,z) \rvert} =k_A(x,z_b)\exp\left[\frac{2\pi(z-z_b)}{L}\right],
    \end{equation}
    \begin{equation}
        b_{y 0}(x,z) = -  \sqrt{\mu_0 \bar{\rho}}\;\text{sgn}(B_{0 z}) v_0(x,z),
    \end{equation}
\end{subequations}
\rev{similar to before. The amplitude function is found at the next order in $\varepsilon$:}
\begin{equation}\label{eqn:phasedampingamp}
    v_0(x,z)=v_b(x)\exp\left\{-\frac{\eta}{2\omega}\int_{z_b}^z k_A\left[k_A^2+ \left(\partial_x\theta_A\right)^2\right]\mathrm{d}z'\right\}.
\end{equation}

\rev{To characterize the rate of damping, we define a damping distance $\lambda$ as the vertical distance from the height at which the AWs are forced ($z=z_b$) to the height at which their amplitude has decreased by a factor of $1/e$ (i.e., $v_0(x,z_b+\lambda)=v_b(x)/e$). Using (\ref{eqn:kA}) and (\ref{eqn:phasedampingamp}), it may be shown that for our assumption of exponentially-decaying magnetic fields, $\lambda = (L\ln\zeta)/(2\pi)$, where $\zeta$ is the only real root of}
\begin{equation}\label{eqn:cubiczeta}
    \alpha \zeta^3 - \zeta^2 + \zeta - \alpha - {\beta} \textit{Lu}_L = 0,
\end{equation}
\rev{with $\alpha(x) = \frac{1}{3}(4\pi^2 a^2+1)$, $\beta(x) = 2a^2[2\pi/ (L k_A(x,z_b))]^3 $, and $a(x) = k_A/(L \partial_x k_A)$.}

\rev{As $\textit{Lu}_L \to \infty$, the damping distance scales as}
\begin{equation}\label{eqn:asympdampingdistance}
    {\lambda} \sim \frac{L}{2\pi} \ln\left[\frac{2\pi}{{k}_A({x},{z}_b) L}\left(\frac{6 \textit{Lu}_L}{4\pi^2 + a^{-2}}\right)^{1/3}\right].
\end{equation}
\rev{Because our magnetic field decays exponentially with height, we find that the damping length scales like $\ln(\textit{Lu}_L^{1/3})$, which is much shorter than the classical $\textit{Lu}_L^{1/3}$ scaling for the damping length $\bar{\lambda}$ in the case of a vertically-invariant background field:}
\begin{equation}
    \bar{\lambda} \sim \frac{1}{k_A(x)}\left(6 a^2 \textit{Lu}_L\right)^{1/3}
\end{equation}
\rev{\citep[][Equation 20]{heyvaerts_coronal_1983}. As AWs propagate up into regions of lower magnetic field strength, they develop shorter vertical wavelengths (higher $k_A$). As a result, the AWs become phase-mixed over shorter propagation distances and develop finer horizontal scales over which diffusion can act more efficiently \citep{de_moortel_phase_2000}.}

\rev{Diffusion sets the minimum horizontal scale $\delta$ produced by phase mixing, which may be estimated from the horizontal gradient of $v$ at $z=z_b+\lambda$. Since $v^{-1}\partial_x v \sim i \int_{z_b}^z\partial_x k_A(x,z') \mathrm{d}z'$ when $z=O(L)$,}
\begin{equation}
    \delta \sim \left[\int_{z_b}^{z_b+\lambda}\partial_x k_A(x,z) \mathrm{d}z\right]^{-1}.
\end{equation}
\rev{As $\textit{Lu}_L \to \infty$, the horizontal diffusive scale follows}
\begin{equation}\label{eqn:diffusivescale}
    \delta \sim \hat{\delta} = L \left[\frac{6 \textit{Lu}_L}{a(4\pi^2 a^2 + 1)}\right]^{-1/3},
\end{equation}
\rev{for our assumption of an exponentially-decaying magnetic field \citep[a similar scaling is obtained for the width of AW resonant layers in][]{kappraff_resistive_1977,hollweg_resonance_1988,poedts_efficiency_1990}.}

\rev{For comparison to our simulations, we predict the damping distance $\lambda$ (by solving Equation \ref{eqn:cubiczeta}) and the diffusive scale $\hat{\delta}$ (from Equation \ref{eqn:diffusivescale}) with $B_{0 z} = \cos(2\pi x/L)\exp(-2\pi z/L)$. The AW starting height is set to $z_b = 0.09 L$ (the height at which AWs first appear in the simulations), and the length-scale predictions are evaluated at $x/L = 1/3$ (away from singularities in $k_A$ and zeros of $\partial_x k_A$). For IVP I, $\lambda \approx 0.04 L$; Figure \ref{fig:casei}(\textit{f}) shows that the AWs are indeed strongly damped when they reach $z_b + \lambda \approx 0.13 L$. The lower diffusivity in IVP II results in a longer damping distance ($z_b + \lambda \approx 0.21 L$) such that the AWs are still faintly visible at the forcing height in Figure \ref{fig:caseii}(\textit{f}).}

\rev{The smallest horizontal AW length-scales decrease by half between IVP I and IVP II, and they are roughly 10 times larger than $\hat{\delta}$ in both simulations ($\hat{\delta}/L \approx 5.9\times10^{-4}$ in IVP I and $\hat{\delta}/L \approx 2.7\times10^{-4}$ in IVP II), following the $\textit{Lu}^{-1/3}$ scaling prediction. The resistive scale $\hat{\delta}$ may be used to predict the time $\tau_e$ required for the simulations to equilibrate: $\omega \tau_e \sim \omega \hat{\delta}^2/\eta \propto \textit{Lu}_L^{1/3}$. Indeed, we find that the 10-fold increase in $\textit{Lu}_L$ from IVP I to IVP II roughly doubles ($10^{1/3} \approx 2$) the time for the total energy to saturate.}

\rev{The AW length-scales observed in our simulations can be scaled to stellar parameters using the asymptotic predictions in (\ref{eqn:cubiczeta}) and (\ref{eqn:diffusivescale}). In the cores of RGB stars, the viscosity is likely larger than the magnetic diffusivity \citep{skoutnev_zones_2025}, and so the role of $\textit{Lu}_L$ is replaced by $\textit{Re}_L = \omega L^2/\nu \approx 10^{16}$ to $10^{21}$ \citep[using dimensional values from ][]{denissenkov_numerical_2010,griffiths_magneto-rotational_2022,montalban_testing_2013,stello_prevalence_2016}, where $L$ is the depth of the core, $\nu$ is the kinematic viscosity, and $\omega$ is the frequency of depressed dipole modes. For simplicity, we approximate the core magnetic field with the $\Gamma =0.1$ ``dipolar'' field used in our simulations, which decays exponentially over the depth of the core $L$. Then, the diffusive scale of the AWs is predicted as $10 \hat{\delta}/L \approx 10^{-7}$ to $10^{-5}$ (roughly 100 m to 5 km using a core radius of $L\approx 0.6 R_\odot$).}

\rev{Assuming again that the background field decays exponentially over the depth of the core $L$, the AW damping distance is $\lambda \approx0.92 L$ to $1.5 L$, which would suggest that AWs could propagate all the way to the base of the convective envelope with little dissipation. However, the damping distance can be much shorter if the background magnetic field experiences faster-than-exponential decay with height. At the outer extent of a fossil field confined to the radiative zone \citep[e.g.,][]{skoutnev_magnetic_2025}, the AW wavenumber $k_A$ diverges and the local damping distance $(z-z_b)/\ln(v_b/v_0)$ tends to zero. Regardless of their damping rates, we find no evidence that the phase-mixed AWs should develop coherent large-scale structures and convert back to up-going IGWs.}

In the next section, \rev{we neglect the AWs and focus on the large-scale features in our simulations, which we characterize} as wave modes found via a WKB approximation in the vertical direction. The following WKB analysis elucidates the mechanism of wave conversion and explains why the large scale modes have the same behavior in IVP I and II.

%% file: 4-wkb.tex
In this Section, we use a WKB analysis to derive an expression for the large-scale oscillation modes in the limit of small $\textit{Fr}$. The approximation takes advantage of the separation between the short vertical scale of the oscillations ($O(\textit{Fr} L)$) and the slow variation of the background magnetic field ($O(L)$). We approximate each field as, e.g.,
\begin{equation}\label{eqn:waveansatz}
    p \sim A(z) p_0(x,z)\exp\left(i\theta(z) + i k_y y - i \omega t\right),
\end{equation}
where $k_y = 2\pi/L$ and $\omega$ are fixed by the forcing. The local scale of oscillations is described by the vertical wavenumber $k_z(z) = \partial_z \theta$. Crucially, the WKB approximation assumes that $\theta(z)$ varies rapidly with $z$ over $O(\textit{Fr} L)$ scales while $k_z(z)$, $p_0(x,z)$, and $A(z)$ vary slowly with $z$ on the system scale $L$. The amplitude function $A(z)$ is common to all the fields ($\boldsymbol{u}, p, \rho, \boldsymbol{b}$), and is found by imposing a solvability condition on the next-to-leading order equations. The details of the WKB procedure and amplitude equation (\ref{eqn:amp}) may be found in Appendix \ref{app:wkb}.

\begin{figure*}[ht!]
    \centering
    \includegraphics[width=\linewidth]{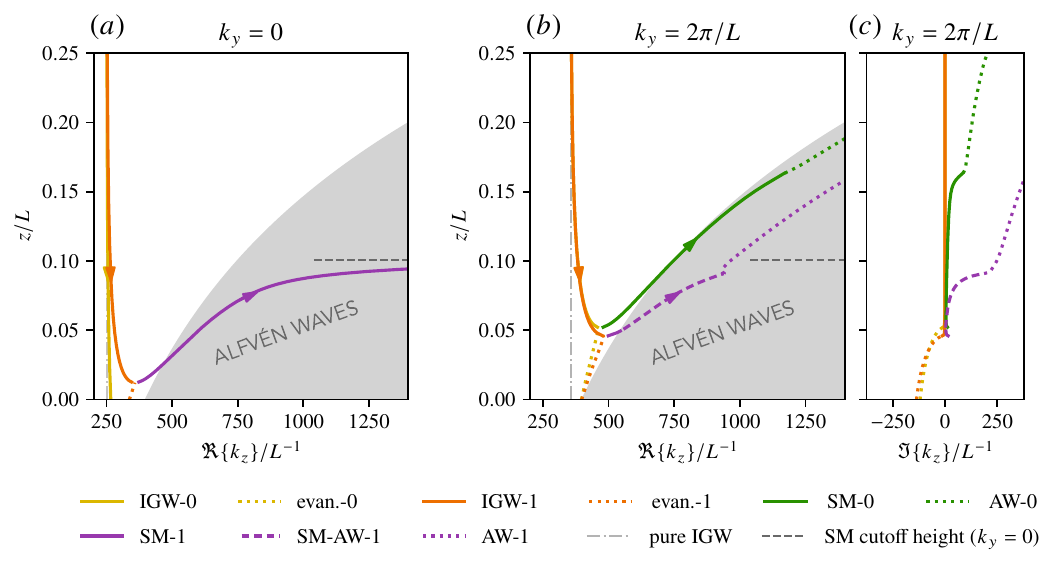}
    \caption{(\textit{a}) The real part of the local vertical wavenumber $k_z(z)$ versus height $z$ in the axisymmetric WKB problem ($k_y=0$, $\Gamma = 0.1$) in Section \ref{sec:wkb:axi}. Solid curves correspond to five discrete wave modes (IGW-0, IGW-1, evan.-0, evan.-1, SM-1) while the gray shaded region indicates the continuous spectrum of AWs (\ref{eqn:alfvenbdry}) at each height. Arrows indicate the direction of the WKB group velocity, computed using (\ref{eqn:groupvel}). IGW modes closely follow the wavenumber prediction for pure IGWs (vertical gray dash-dotted line) based on (\ref{eqn:lowFrigwdisprel}). A SM cutoff height (horizontal gray dashed line) at $z/L = \ln(6\pi \Gamma)/(2\pi) \approx 0.101$ bounds the upward path of SM-1. Note that two SM modes may exist above $z=0$ for higher values of $\Gamma$ (cf. Figure 5 of \citealt{lecoanet_conversion_2017}). (\textit{b}) The real part of $k_z(z)$ in the resistive nonaxisymmetric WKB problem ($k_y=2\pi/L$, $\Gamma = 0.1$) in Section \ref{sec:wkb:nonaxi}. The five waves present in the axisymmetric case are now accompanied by a sine-parity SM wave (SM-0), two AWs (AW-0 and AW-1), and a mixed SM-AW mode (SM-AW-1). The latter refracts as it approaches the same SM cutoff height (dashed gray line) as in panel \textit{a}. (\textit{c}) The imaginary part of the WKB wavenumber $k_z(z)$ for $k_y = 2\pi/L$ and $\Gamma = 0.1$. For the up-going SM, SM-AW, and AW modes, $\Im\{k_z\} > 0$ implies locally exponential damping and is due to Ohmic dissipation.}
    \label{fig:evals}
\end{figure*}

We assume that the strength of the magnetic field is commensurate with the degree of stratification such that $\Gamma$ remains $O(1)$ as $\textit{Fr} \to 0$. In this distinguished limit, we show in Appendix \ref{app:wkb} that the WKB approximation reduces the linear equations (\ref{eqn:magbouss}) to a one-dimensional generalized eigenvalue problem in $x$ that depends parametrically on $z$ through $\boldsymbol{B}_0(x,z)$. Ignoring the effects of magnetic diffusion, the problem can be reduced to a single equation for the pressure eigenfunction $p_0(x,z)$ and eigenvalue $k_z^2(z)$:
\begin{equation}\label{eqn:idealpressure}
    \frac{k_z^2}{N^2} p_0 + \hat{\boldsymbol{\nabla}} \boldsymbol{\cdot}\left(\frac{ \hat{\boldsymbol{\nabla}} p_0}{\omega^2 - v_{A z}^2 k_z^2}\right) = 0,
\end{equation}
where $\hat{\boldsymbol{\nabla}} = \boldsymbol{e}_x \partial_x +  \boldsymbol{e}_y i k_y$ is the transformed horizontal gradient operator and
\begin{equation}
    v_{A z}(x,z) = \frac{\boldsymbol{B}_0 \boldsymbol{\cdot} \boldsymbol{e}_z }{\sqrt{\mu_0 \bar{\rho}}}
\end{equation}
is the local Alfv\'en speed. (Note that this eigenproblem can be solved independently at each height $z$ in the domain). The pressure equation (\ref{eqn:idealpressure}) is the Cartesian analog of the spherical eigenvalue equation in \citet[Equation 12]{rui_gravity_2023}. 

If $v_{A z} \equiv 0$, then (\ref{eqn:idealpressure}) reduces to the (pure) IGW dispersion relation in the $\textit{Fr} \to 0$ limit (cf. Equation \ref{eqn:igwdisprel}):
\begin{equation}\label{eqn:lowFrigwdisprel}
    \frac{k_z^2}{N^2} - \frac{k_x^2 + k_y^2}{\omega^2} = 0,
\end{equation}
for which the eigenfunctions are Fourier modes $p_0 = \exp(i k_x x)$ and the eigenvalues ($k_z^2$) are independent of $z$. In our model, $v_{A z} \neq 0$ and $\partial_z v_{A z} \neq 0$, so both the eigenfunctions $p_0(x,z)$ and the vertical wavenumber $k_z(z)$ must vary (slowly) with $z$.

\subsection{Axisymmetric modes}\label{sec:wkb:axi}

Figure \ref{fig:evals}\textit{a} shows the real part of the vertical wavenumber $k_z$ versus height $z$ for the axisymmetric ($k_y = 0$) case (similar to the constant $N$ case analyzed by \citealt{lecoanet_conversion_2017}) \rev{with $\boldsymbol{B}_0$ given by (\ref{eqn:backgroundfield})}. The eigenvalues $k_z^2$ are found as described in Appendix \ref{app:evp:idealaxi}. Only the wavenumbers of ``dipolar'' modes (i.e., those for which $\lvert p_0\rvert$ has two zeros on $x \in [0,L)$ at each $z$) are included. Arrows are oriented along the wavenumber branches according to the sign of the WKB (vertical) group velocity, which is computed using (\ref{eqn:groupvel}).

Figure \ref{fig:evals}\textit{a} shows 2 overlapping branches of $k_z$ corresponding to the sine-parity IGW (IGW-0, yellow curve) and the cosine-parity IGW (IGW-1, orange curve), respectively. As before, we refer to the modes by the parity of the associated latitudinal flow, $u_0$. Both IGW-0 and IGW-1 branches closely follow the prediction for pure IGWs ($k_z = \textit{Fr}^{-1}\sqrt{k_x^2 + k_y^2}$, vertical gray dash-dotted line) near the top of the domain, but refract to higher wavenumber as they propagate down into regions of greater magnetic field strength. 

At $z/L \approx 0.0125$, IGW-1 reaches a turning point, below which it leaves an an evanescent tail (evan.-1, dotted orange curve in Figure \ref{fig:evals}\textit{a}) with $\Im\{k_z\} < 0$. Note that for the given magnetic field strength ($\Gamma = 0.1$), the corresponding turning point for IGW-0 is outside the domain at $z/L = \ln(2\pi \Gamma/1.08)/(2\pi) \approx -0.0862$ \citep[see Table 2 in][]{lecoanet_conversion_2017}.

Above the IGW-1 turning point, there is a connecting wavenumber branch corresponding to an up-going cosine-parity SM mode (SM-1, purple curve in panel \textit{a}). Since the restoring mechanism for SM waves is dominated by the Lorentz force rather than buoyancy (as for IGWs), the properties of SM-1 are intrinsically tied to the background magnetic field. The wave refracts again as it propagates up to regions of lower magnetic field strength. The upward path of SM-1 is bounded by a magnetic cutoff height (Figure \ref{fig:evals}\textit{a}, horizontal gray dashed line) at $z/L = \ln(6\pi \Gamma)/(2\pi) \approx 0.101$ \citep{lecoanet_conversion_2017}, near which $k_z$ becomes infinitely large and the wave damps for arbitrarily small diffusivity. 

\cite{rui_gravity_2023} found an important difference between these behaviors (in a Cartesian domain) and the corresponding behavior of IGWs and SM waves in spherical geometry with a dipolar background field: for axisymmetric dipole modes ($\ell=1$, $m=0$), they find that the IGW approaches a magnetic cutoff height from above and never reaches a turning point. However, they show that $\ell > 1$, $m=0$ modes exhibit similar IGW$\rightarrow$SM turning points as in the Cartesian analysis of \cite{lecoanet_conversion_2017} and Figure \ref{fig:evals}(\textit{a}). Moreover, we expect the inconsistent behavior for the $\ell=1$, $m=0$ case to have little bearing on nonaxisymmetric dipole modes since \cite{rui_gravity_2023} also find IGW$\rightarrow$SM turning points for $\ell = |m|=1$, as discussed in Section \ref{sec:wkb:nonaxi}.

As \cite{rui_gravity_2023} further note, the ideal magnetogravity eigenproblem (\ref{eqn:idealpressure}) has an internal singularity at
\begin{equation}\label{eqn:AWdisp}
    \omega^2 - v_{A z}^2 k_z^2 = 0,
\end{equation}
which is also present in the polarization relation for $v_0$ (the azimuthal velocity eigenfunction):
\begin{equation}\label{eqn:vpolrel}
    v_0 = \frac{k_y \omega p_0}{\bar{\rho}(\omega^2 - v_{A z}^2 k_z^2)}.
\end{equation}
Since the local Alfv\'en speed $v_{A z}$ varies continuously with $x$, this singularity is associated with a continuous spectrum of AW modes (gray shaded region in Figure \ref{fig:evals}\textit{a}) with
\begin{equation}\label{eqn:alfvenbdry}
    \lvert k_z \rvert \geq k_{A B}(z) = \frac{\omega}{\max_x \lvert v_{A z}(x,z)\rvert} =\frac{1}{\Gamma l_z} \exp\left(\frac{2\pi z}{L}\right)
\end{equation}
satisfying (\ref{eqn:AWdisp}) somewhere in the domain. The separation between the IGW branches ($k_z \sim l_z^{-1}$) and the Alfv\'en continuum ($k_z \sim \Gamma^{-1} l_z^{-1}$) decreases inversely with the magneto-gravity ratio $\Gamma$.

When $k_y = 0$, the singularity in (\ref{eqn:idealpressure}) may be removed after integrating the equation with respect to $x$. Then, (\ref{eqn:idealpressure}) and (\ref{eqn:vpolrel}) yield decoupled dispersion relations involving the meridional ($x$, $z$) streamfunction $\psi_0 = (\omega k_z\int_0^x p_0 \mathrm{d}x')/(\bar{\rho} N^2)$ and the azimuthal flow $v_0$, respectively:
\begin{subequations}
    \begin{equation}
        \omega^2 k_z^2 \psi_0 - v_{A z}^2 k_z^4 \psi_0 + N^2 \partial_x^2 \psi_0 = 0,
    \end{equation}
    \begin{equation}
        (\omega^2 - v_{A z}^2 k_z^2)v_0 = 0.
    \end{equation}
\end{subequations}
In this case, each AW eigenfunction $v_0$ may be expressed as the sum of Dirac delta functions
\begin{equation}\label{eqn:AWaxisymmode}
    v_0 = \frac{1}{N_c}\sum_{n=1}^{N_c} \delta(x-x_{c,n})
\end{equation}
centered at critical latitudes $x_c$ where $\omega^2 - v_{A z}^2 k_z^2 = 0$. Physically, (\ref{eqn:AWaxisymmode}) and (\ref{eqn:alfvenbdry}) mean that each magnetic field line oscillates in the azimuthal direction with a slightly different vertical wavenumber than its neighbors when $k_y = 0$ \citep[e.g.,][]{heyvaerts_coronal_1983,Loi2017}. As shown in Section \ref{sec:results} (see Figure \ref{fig:phasemixing}), large-scale azimuthal motions can project onto a continuous spectrum of singular Alfv\'en eigenmodes and produce fine horizontal scales via phase mixing.

This effect is unimportant in the $k_y = 0$ case \citep{lecoanet_conversion_2017} since the IGWs forced at the top of the domain are decoupled from azimuthal motions $v$. However, in the $k_y \neq 0$ case considered below, the IGWs have nonzero $v$. To see this, first observe that the perturbations in density ($\rho_0 \neq 0$) associated with IGWs must be accompanied by pressure fluctuations (the $z$ component of the momentum equation, \ref{eqn:zmomorder1}, implies that $p_0 \neq 0$ if $\rho_0 \neq 0$). Then, the $y$ components of the induction and momentum equations (combined as Equation \ref{eqn:vpolrel}) imply that these pressure fluctuations induce azimuthal motions ($v_0 \neq 0$) if $k_y \neq 0$. The consequent coupling of the discrete modes to the Alfv\'en continuum results in the conversion to AWs and to a mixed SM-AW mode as discussed below.

\subsection{Nonaxisymmetric modes}\label{sec:wkb:nonaxi}

The AWs and mixed SM-AW mode have features with fine scales in $x$ associated with the singularity in (\ref{eqn:vpolrel}), as shown later in this section. To regularize these features with minimal influence on the remaining large-scale modes, we introduce anisotropic diffusion to the induction equation:
\begin{equation}\label{eqn:anisoinduction}
    \partial_t \boldsymbol{b} = \boldsymbol{\nabla}\times(\boldsymbol{u}\times\boldsymbol{B}_0) +  \eta_x \partial_x^2 \boldsymbol{b} + \eta (\partial_y^2 + \partial_z^2) \boldsymbol{b},
\end{equation}
with higher diffusivity in the $x$ direction ($\textit{Lu}_x = \omega l_z^2/\eta_x = 12.5$) than in $y$ or $z$ ($\textit{Lu} = 6.25 \times 10^4$). We retain the leading order anisotropic diffusion terms in the reduced induction equation:
\begin{equation}\label{eqn:anisoinduction-reduced}
	-i \omega \boldsymbol{b}_{0 H} = i k_z B_{0 z} \boldsymbol{u}_{0 H} +  \eta_x \partial_x^2 \boldsymbol{b}_{0 H} - \eta (k_y^2 + k_z^2) \boldsymbol{b}_{0 H},
\end{equation}
where $\boldsymbol{u}_{0 H} = u_0 \boldsymbol{e}_x + v_0 \boldsymbol{e}_y$ and $\boldsymbol{b}_{0 H} = b_{0 x}\boldsymbol{e}_x + b_{0 y} \boldsymbol{e}_y$. (Terms involving $w_0$ and $b_{0 z}$ are $O(\textit{Fr})$ smaller and thus neglected). The full set of (resistive) reduced equations are derived in Appendix \ref{app:wkb}. Appendix \ref{app:evp:resnonaxi} presents the reduced system as a generalized eigenvalue problem and details the procedure for finding the nonaxisymmetric ($k_y = 2\pi/L$) resistive modes.

Figure \ref{fig:evals}\textit{b} plots the real part of $k_z$ for the $k_y = 2\pi/L$ resistive modes. In the presence of diffusion, the ideal Alfv\'en continuum is replaced by discrete wavenumber branches \citep[e.g.,][]{hoven_magnetar_2011}. However, these branches are so closely spaced for $\textit{Lu}_x =12.5$ and $\textit{Lu} = 6.25 \times 10^4$ that we plot the ideal Alfv\'en continuum (gray shaded region) for simplicity. As in the $k_y = 0$ case, IGW-0 and IGW-1 (yellow and orange curves, respectively) closely follow the pure IGW wavenumber (gray vertical dash-dotted line) near the top of the domain. The two parities exhibit different behavior as they interact with background magnetic field.

There is a turning point at $z/L \approx 0.0461$, where IGW-1 meets the SM-1 branch (purple curve in Figure \ref{fig:evals}\textit{b}) and leaves behind an evanescent tail (evan.-1, dotted orange curve), as before. However, beyond the Alfv\'en wavenumber boundary (i.e., when $\Re\{k_z (z)\} \ge k_{A B}(z)$, see Equation \ref{eqn:alfvenbdry}), SM-1 becomes a mixed SM-AW mode (SM-AW-1, dashed purple curve). The $v_0$ eigenmodes for SM-AW-1 resemble SM-1 apart from sharp features at critical latitudes $x_c$ where the mode resonates with the Alfv\'en continuum. Despite these resonances, SM-AW-1 behaves similarly to SM-1, refracting as it approaches the $k_y = 0$ SM cutoff height (horizontal gray dashed line). The increase of the imaginary part of $k_z$ with height (see Figure \ref{fig:evals}\textit{c}, dashed purple curve) shows that SM-AW-1 decays faster than exponentially as it approaches the cutoff height.

This behavior is consistent with the results of \cite{rui_gravity_2023} in spherical geometry. For nonaxisymmetric dipole modes ($\ell = \lvert m \rvert = 1$), they identify a turning point between distinct IGW and SM branches. Further, they find that the SM eigenmodes develop sharp features at critical latitudes when the associated radial wavenumber satisfies the Alfv\'en dispersion relation somewhere in the domain. In this work, we characterize these modes as mixed SM-AW modes and distinguish them from AWs that do not directly connect to SM branches.

The mixed SM-AWs have large-scale features away from the critical latitudes and may be resolved with vertical diffusion ($\eta \partial_z^2 \boldsymbol{b} \rightarrow -\eta k_z^2 \boldsymbol{b}_0$) alone, as found by \cite{lecoanet_asteroseismic_2022} and \cite{rui_gravity_2023}. In contrast, the AWs are dominated by fine scale oscillations and are only resolved when the horizontal diffusion operator ($\eta_x \partial_x^2 \boldsymbol{b}_0$) is explicitly included in the WKB induction equation (\ref{eqn:inductionorder1}).

These AWs have wavenumbers with large imaginary part for the values of the diffusivities used here. Note that $\Im\{k_z\}$ depends mainly on $\eta_x$ here, since $\eta_x \gg \eta$; we expect $\Im\{k_z\}/\Re\{k_z\}$ would be much smaller if $\textit{Lu}_x = \textit{Lu} = 6.25 \times 10^4$. Near the cutoff height, SM-AW-1 connects to one such branch of evanescent AWs (AW-1, dotted purple curve). Unlike SM-AW-1, the azimuthal velocity eigenmode $v_0$ for AW-1 grows faster with height than $u_0$; the AW transfers energy from latitudinal motions to azimuthal ones (though these still ultimately damp since $v$ depends on the product $v_0 \exp(i k_z z)$ where $\Im\{k_z\} > 0$).

The wave interactions for the opposite parity are slightly different. At $z/L \approx 0.0522$, IGW-0 refracts into a sine-parity SM branch (SM-0, green curve in Figure \ref{fig:evals}\textit{b}) and leaves behind an evanescent tail (evan.-0, dotted yellow curve). SM-0 refracts further as it propagates upwards, following the Alfv\'en wavenumber boundary $k_z \approx k_{A B}(z)$. Near $z/L \approx 0.163$, the SM-0 branch enters the Alfv\'en continuum (i.e., $\Re\{k_z(z)\} > k_{A B}(z)$) and connects to an evanescent AW branch (AW-0, dotted green curve).

\subsection{Comparison between WKB theory and simulation}
\begin{figure*}[ht!]
    \centering
    \includegraphics[width=\linewidth]{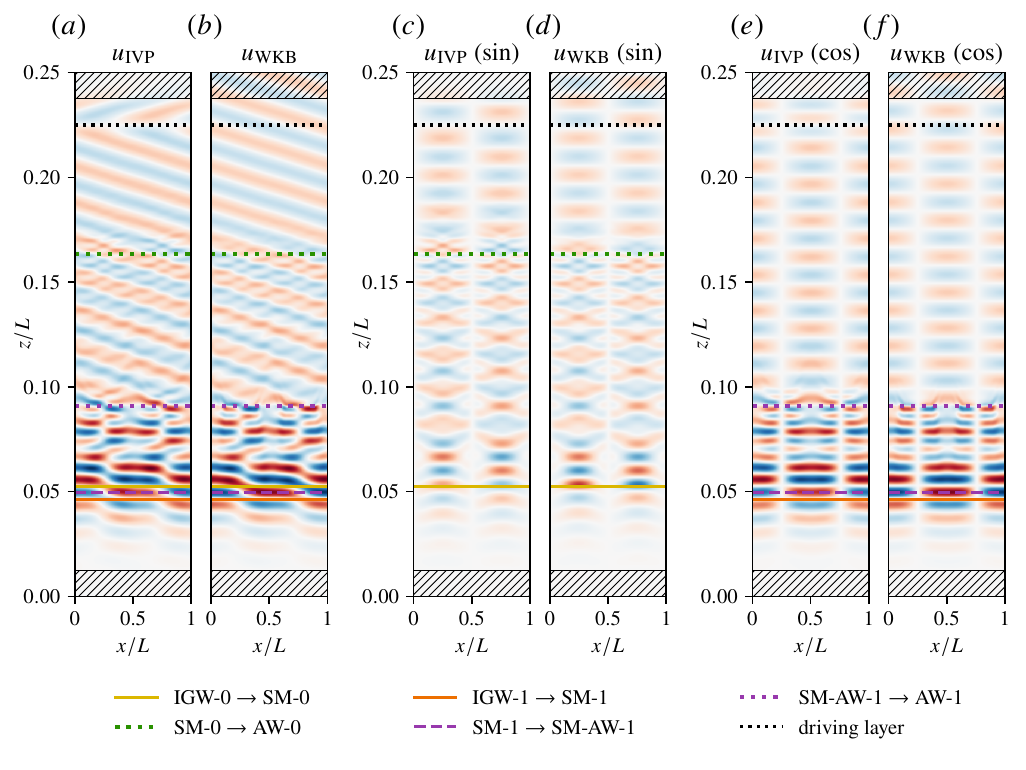}
    \caption{(\textit{a}) Snapshot of the latitudinal velocity $u_{\text{IVP}}$ from IVP III ($\textit{Lu}_x = 12.5$, $\textit{Lu} = 6.25 \times 10^4$, $\textit{Fr} = 0.025$, $\Gamma =0.1$) after the simulation has equilibrated. Horizontal lines correspond to the wave mode transition heights predicted by Figure \ref{fig:evals}. (\textit{b}) The latitudinal velocity $u_{\text{WKB}}$ predicted from the WKB approximation. There is agreement with $u_{\text{IVP}}$ below the driving height and away from the turning points. (\textit{c})–(\textit{d}) Sine-parity components of $u_{\text{IVP}}$ and $u_{\text{WKB}}$, respectively. (\textit{e})–(\textit{f}) Cosine-parity components of $u_{\text{IVP}}$ and $u_{\text{WKB}}$, respectively.}
    \label{fig:uwkb}
\end{figure*}
\begin{figure*}[ht!]
    \centering
    \includegraphics[width=\linewidth]{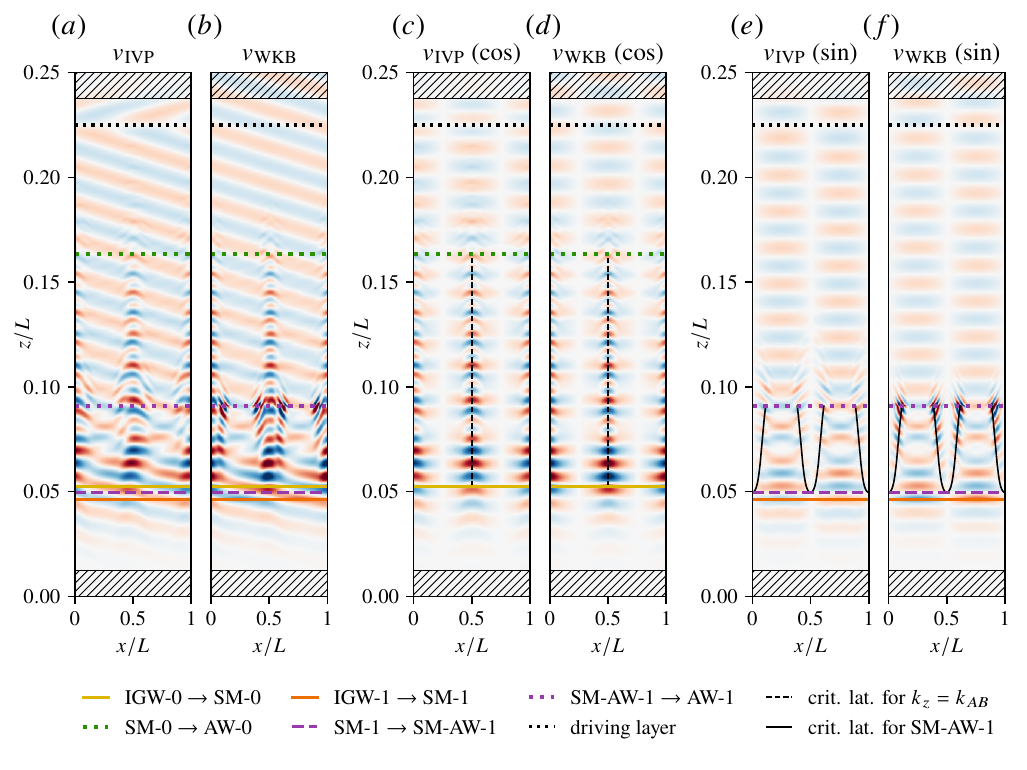}
    \caption{Redux of Figure \ref{fig:uwkb}, showing the azimuthal velocity $v$ instead of the latitudinal velocity $u$. (\textit{a})–(\textit{b})  The equilibrated azimuthal velocity from IVP III ($v_{\text{IVP}}$) and the corresponding WKB approximation ($v_{\text{WKB}}$). (\textit{c})–(\textit{d}) Cosine-parity components of $v_{\text{IVP}}$ and $v_{\text{WKB}}$, respectively. As SM-0's vertical wavenumber draws nearer to the Alfv\'en continuum, the wave develops sharper ridge-like features centered at critical latitudes $x_c/L = 0,0.5,1$ (vertical dashed lines) associated with the Alfv\'en wavenumber boundary $k_z = k_{A B}(z)$. (\textit{e})–(\textit{f}) Sine-parity components of $v_{\text{IVP}}$ and $v_{\text{WKB}}$, respectively. The mixed SM-AW mode (SM-AW-1) develops sharp shingle-shaped features near critical latitudes (solid black curves) at which the WKB wavenumber $k_z$ satisfies the ideal Alfv\'en dispersion relation (\ref{eqn:AWdisp}).}
    \label{fig:vwkb}
\end{figure*}

The reduced eigenvalue problem suggests that IGWs of both parity convert to decaying magnetohydrodynamic waves. To confirm this, we compare a WKB solution composed of the nine dipolar modes discussed above (IGW-0, IGW-1, evan.-0, evan.-1, SM-0, SM-1, SM-AW-1, AW-0, AW-1) to a third numerical simulation, IVP III, that uses anisotropic magnetic diffusion. In IVP III, we solve (\ref{eqn:anisoinduction}) instead of (\ref{eqn:lininduction}), with $\textit{Lu}_x = 12.5$ and $\textit{Lu} = 6.25 \times 10^4$ to match the nonaxisymmetric eigenvalue problem.

Figure \ref{fig:uwkb} compares the latitudinal velocity component of IVP III in the equilibrated state ($u_{\text{IVP}}$, panel \textit{a}) to the WKB solution ($u_{\text{WKB}}$, panel \textit{b}). A supplementary video available at \rev{\url{https://doi.org/10.5281/zenodo.18357092}} shows the transient behavior of IVP III (details under Supplementary Materials). Because the WKB solution is only determined up to an arbitrary phase and amplitude, comparing it to the IVP solution requires a few steps, the full details of which are in Appendix \ref{app:wkbsoln}. The overall amplitude and phase of $u_{\text{WKB}}$ is determined by normalizing the WKB solution to $u_{\text{IVP}}$ at an arbitrary point well above the cutoff heights and away from the zeros of the eigenmodes: $(x/L, z/L) = (1/3, 0.175)$. Crucially, we assume that there is total conversion from IGWs to SM waves at the turning points $z_{t,\text{IGW-0}} \approx 0.0522 L$ and $z_{t,\text{IGW-1}} \approx 0.0461 L$ for the sine- and cosine-parity waves, respectively. Further, we assume that the difference in phase $\theta$ across each turning point is $-\pi/2$, as in the axisymmetric case \citep{lecoanet_conversion_2017}. This assumption, along with that of total conversion between IGWs and SM waves, is supported by the excellent match between $u_{\text{IVP}}$ and $u_{\text{WKB}}$ outside of the damping regions (hatched areas) and below the forcing height (black dotted line).

To enable detailed comparison between the simulation and WKB theory, Figure \ref{fig:uwkb}\textit{c},\textit{d} shows the sine-parity components of $u_{\text{IVP}}$ and $u_{\text{WKB}}$, the latter consisting only of IGW-0, evan.-0, SM-0, and AW-0 modes. The numerical simulation agrees with the results of the WKB analysis assuming total conversion from IGW-0 to SM-0 at the turning point $z = z_{t,\text{IGW-0}}$ (yellow line). SM-0 then damps as it approaches a magnetic cutoff height near $z = z_{t,\text{AW-0}} \approx 0.163 L$ (the diamond-shaped features in Figure \ref{fig:uwkb}\textit{c},\textit{d} vanish near the dotted green line).

Figure \ref{fig:uwkb}\textit{e},\textit{f} shows just the cosine-parity components of $u_{\text{IVP}}$ and $u_{\text{WKB}}$, the latter consisting only of IGW-1, evan.-1, SM-1, SM-AW-1, and AW-1 modes. Again, the match between numerical simulation and WKB theory supports total conversion from IGW to SM wave at the turning point $z=z_{t,\text{IGW-1}}$ (orange line). SM-1 then crosses the Alfv\'en wavenumber boundary (at the dashed purple line) and completely converts to a mixed SM-AW mode (SM-AW-1). SM-AW-1 damps rapidly as it approaches $z = z_{t,\text{AW-1}} \approx 0.091 L$ (the lenticular features in Figure \ref{fig:uwkb}\textit{e},\textit{f} vanish near the dotted purple line).

Figure \ref{fig:vwkb} compares the azimuthal velocity component of IVP III in the equilibrated state ($v_{\text{IVP}}$, panel \textit{a}) to the WKB solution ($v_{\text{WKB}}$, panel \textit{b}). The latter is constructed using the same amplitude and phase shifts as $u_{\text{WKB}}$. The agreement between WKB theory and simulation is not as good for $v$ as it is for $u$, owing to the strong influence of the Alfv\'en continuum on $v$ (discussed below).

\begin{figure*}[ht!]
    \centering
    \includegraphics[width=\linewidth]{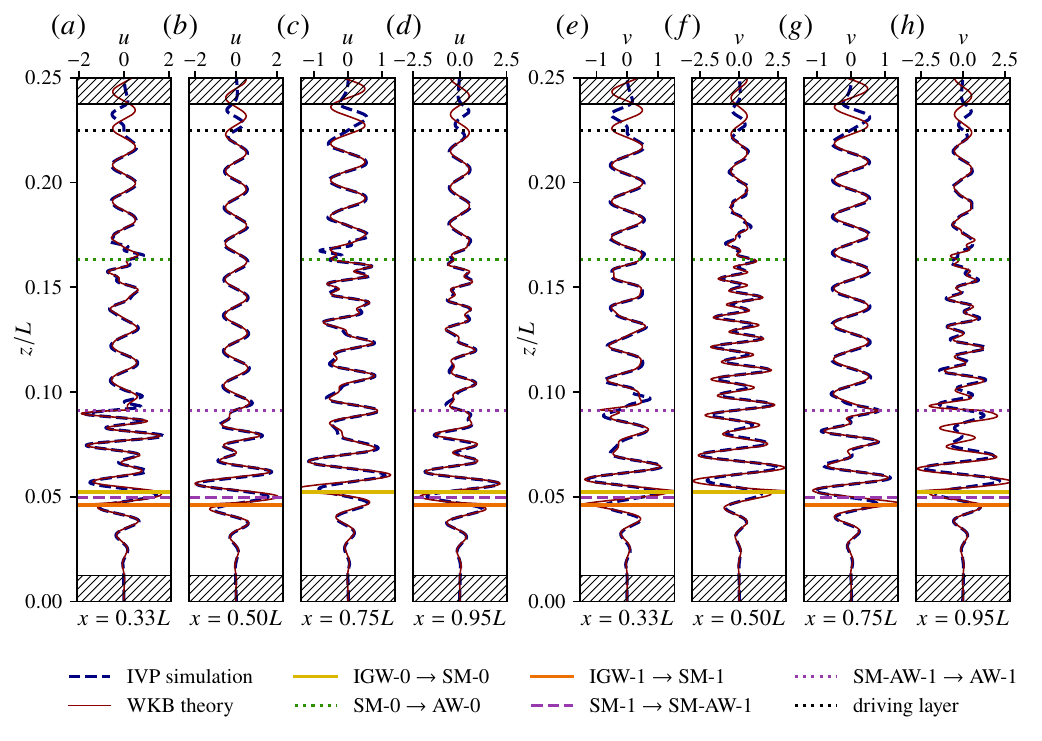}
    \caption{(\textit{a})–(\textit{d}) Comparison of the equilibrated latitudinal velocity from IVP III ($u_{\text{IVP}}$) and the corresponding WKB approximation ($u_{\text{WKB}}$) at different latitudes $x$. We fit the WKB solution to $u_{\text{IVP}}$ at a single point $(x/L, z/L) = (1/3, 0.175)$ to determine the overall amplitude and phase. (\textit{e})–(\textit{h}) Comparison of the equilibrated azimuthal velocity from IVP III ($v_{\text{IVP}}$) and the corresponding WKB approximation ($v_{\text{WKB}}$).
    }
    \label{fig:profs}
\end{figure*}

Figure \ref{fig:vwkb}\textit{c},\textit{d} shows the cosine components of $v_{\text{IVP}}$ and $v_{\text{WKB}}$. Above the turning point (yellow line), SM-0 develops ridge-like features centered at $x/L = 0,0.5,1$ that sharpen as it approaches the Alfv\'en continuum in wavenumber space. These ridges lie at the critical latitudes associated with the Alfv\'en boundary $k_z(z) = k_{A B}(z)$ (vertical black dashed lines).

Near $z = z_{t,\text{AW-0}}$ (Figure \ref{fig:vwkb}\textit{c},\textit{d}, dotted green line), SM-0 crosses into the Alfv\'en continuum and converts to an evanescent AW (AW-0, faint chevron-shaped features above the dotted green line). Comparison to IVP I (Figure \ref{fig:casei}\textit{e}) and IVP II (Figure \ref{fig:caseii}\textit{e}) shows that the damping rate of the large chevron-shaped features associated with AW-0 depends strongly on the diffusivities. Though phase-mixing is present in IVP I and II, it acts less efficiently on these features since they occur where the local Alfv\'en wavenumber varies most slowly ($x/L = 0, 0.5, 1$). Regardless, the horizontal structure of these AWs is distinct from that of the IGW eigenmodes, and they are not expected to couple to up-going IGWs even in the limit of vanishing diffusivity.

Figure \ref{fig:vwkb}\textit{e},\textit{f} plots the sine components of $v_{\text{IVP}}$ and $v_{\text{WKB}}$. SM-AW-1 develops sharp shingle-like features at critical latitudes (black curves) where the mode resonates with the Alfv\'en continuum (i.e., where $k_z^2 = \omega^2/v_{A z}^2$). Without resistivity, these sharp features correspond to singularities in the azimuthal velocity eigenmode $v_0$. Anisotropic diffusion regularizes these peaks, though it acts slightly differently in the eigenproblem than in the simulation. As a result, the shingles are sharper and have higher amplitude in $v_{\text{WKB}}$ than in $v_{\text{IVP}}$.

It is not entirely surprising that we miss some details of the diffusion-mediated behavior, as it is difficult to include diffusion terms self-consistently in an asymptotic analysis \citep[e.g.,][]{lee_linear_2023}. While we have tried different ways of including diffusion into our asymptotic model, we have not been able to recover the precise behavior of these diffusive modes seen in the IVP. We were only able to obtain eigenfunctions that share the the IVP's broad, smooth peaks at SM-AW-1's critical latitudes by increasing $\eta_x$ such that $\textit{Lu}_x=12.5$. This value of $\eta_x$ is significantly larger than the horizontal diffusivity needed to produce similar features in IVP I ($\textit{Lu}_x = \textit{Lu}= 6.25 \times 10^4$).

Regardless of these differences, the shingle-like features in both IVP III and the WKB solution damp after SM-AW-1 converts to AW-1 (Figure \ref{fig:vwkb}\textit{e},\textit{f}, above the purple dotted line). For lower horizontal diffusivity, energy in SM-AW-1 scatters among AWs that undergo rapid phase mixing with height, as seen in IVP I (Figure \ref{fig:casei}\textit{f}) and IVP II (Figure \ref{fig:caseii}\textit{f}). These AWs damp, and no up-going large-scale features reach the top of the domain.

The behavior of the large-scale wave modes discussed above is supported by the match between IVP III and the WKB solution. To better compare the simulation and theory, Figure \ref{fig:profs} overplots vertical profiles of IVP III (blue dashed curve) and the WKB solution (red curve) at several different latitudes $x/L = 1/3,0.5, 0.75, 0.95$. Panels \textit{a} and \textit{e} plot $u$ and $v$, respectively, at the fitting latitude $x/L = 1/3$. The only major differences between theory and simulation lie above the forcing height (black dotted line) and at the turning points (yellow and orange lines), where the WKB solution is expected to diverge. The excellent agreement between $u_{\text{IVP}}$ and $u_{\text{WKB}}$ persists at $x/L = 0.5, 0.75, 0.95$ (Figure \ref{fig:profs}\textit{b},\textit{c},\textit{d}). 

As discussed above, the quality of fit is slightly diminished for the azimuthal flow $v$. Figure \ref{fig:profs}\textit{f} plots $v$ at $x/L = 0.5$, i.e., only the modes for which $v_0$ has cosine parity (IGW-0, evan.-0, SM-0, AW-0). The ridge-like features of SM-0 (see Figure \ref{fig:vwkb}\textit{c},\textit{d}) are sharper in the WKB theory than in the IVP;  $v_{\text{WKB}}$ has higher amplitude than $v_{\text{IVP}}$ between the yellow and dotted green lines in panel \textit{f}. The WKB solution also overestimates $v$ at $x/L = 0.95$ (panel \textit{h}) between the Alfv\'en boundary intersection height (dashed purple line) and the SM-AW-1 cutoff height (dotted purple line). This mismatch is due to the shingle-like features in SM-AW-1 discussed above (see Figure \ref{fig:vwkb}\textit{e},\textit{f}) that are sharper in the theory than in the simulation.

These differences notwithstanding, the overall agreement between the WKB solution and simulation show that down-going IGWs forced at the top of the domain convert completely to a mix of SM waves and AWs, which develop fine scales and damp. Although we could only explicitly compare the WKB theory to the simulation with anisotropic diffusion (IVP III), the same behavior is observed in the large-scale modes in IVP I and IVP II. Thus, all three simulations support this interpretation.

%% file: 5-conc.tex
This work analyzes the interaction of nonaxisymmetric IGWs with spatially-varying magnetic fields using numerical simulations. We extend the Cartesian model of \cite{lecoanet_conversion_2017} into the third dimension ($y$, azimuth) such that the IGWs and SM waves in their model are accompanied by a continuous spectrum of singular AW modes, as found in the WKB analyses of \cite{lecoanet_asteroseismic_2022} and \cite{rui_gravity_2023} in spherical geometry. While these works are geometrically accurate, they investigate only the eigenvalues and eigenmodes of the reduced equations. By staying in Cartesian geometry, we are able to go beyond \cite{rui_gravity_2023} and obtain an amplitude equation, allowing us to compare a WKB solution directly to numerical simulations and to study the conversion of modes across turning points.

The agreement between our simulations and the WKB theory shows that down-going nonaxisymmetric IGWs convert completely to up-going SM waves (as in the axisymmetric case, \citealt{lecoanet_conversion_2017}). The SM waves then interact with the Alfv\'en continuum in different ways, depending on their parity. The SM wave for which the azimuthal flow vanishes where the vertical background magnetic field component is strongest (SM-1) is at first weakly influenced by the Alfv\'en continuum. SM-1 converts to a mixed SM-AW mode (SM-AW-1), which resembles SM-1 apart from critical layers located at latitudes where the real part of the vertical wavenumber satisfies the Alfv\'en dispersion relation. The SM wave of the opposite parity (SM-0, for which the peaks in azimuthal flow and the vertical component of the background field coincide) is more strongly influenced by the continuous spectrum. Instead of refracting into the Alfv\'en continuum like SM-1, SM-0 propagates higher with a vertical wavenumber that follows the Alfv\'en wavenumber boundary $k_{A B}(z)$ (\ref{eqn:alfvenbdry}). Regardless, both SM-AW-1 and SM-0 approach magnetic cutoff heights near which the latitudinal flow rapidly damps. These results agree with the predictions of \cite{lecoanet_asteroseismic_2022} and \cite{rui_gravity_2023}, who identified similar modes in spherical eigenvalue problems. 

At the magnetic cutoff heights, the SM-AW and SM wave convert to AWs. Phase mixing of AWs produces fine scales that result in efficient damping in the presence of diffusion. Simulations with different magnetic diffusivities ($\textit{Lu} = 6.25 \times 10^4$ in IVP I; $\textit{Lu} = 6.25 \times 10^5$ in IVP II) show that the lower the diffusivity, the further the AWs propagate. Despite these differences, the large-scale modes are the same and agree with the WKB solution in all simulations.

Our results have important implications for the asteroseismic signature of strongly-magnetized stars. In many massive stars, surface convection excites nonradial acoustic waves \citep{houdek_interaction_2015} that couple to IGWs in the radiative interior \citep{Hekker_Mazumdar_2013,takata_asymptotic_2016}. Typically, these IGWs reflect off of the inner region of the core and set up global mixed modes \citep{osaki_nonradial_1975} with predictable fluctuations in surface brightness \citep{Aerts2010}. However, in roughly 20\% of RGB stars, dipole pulsation modes have lower-than-expected amplitudes \citep{mosser_characterization_2012,stello_prevalence_2016}. \cite{lecoanet_conversion_2017} explain the suppression of axisymmetric dipole modes via interactions of down-going IGWs with core magnetic fields above a critical strength.

Importantly, suppressed amplitudes are also observed for nonaxisymmetric dipole modes, and there is no clear pattern of nonaxisymmetric modes being suppressed more or less than axisymmetric modes. The present work shows that these observations could be explained by similar wave interactions with strong magnetic fields. Nonaxisymmetric IGWs convert to a mix of SM and AWs that damp due to the production of fine scales via refraction (in the case of SM waves) and phase-mixing (in the case of AWs). 

It has been suggested that these AWs could travel along closed field loops in the core and convert to up-going IGWs \citep{rui_gravity_2023}. If so, an alternate mechanism would be required to explain the observed suppression of nonaxisymmetric dipole modes. However, we find no evidence of conversion from AWs back to IGWs in our simulations. Though the diffusivities are far lower in RGB stars than in our simulations, we expect efficient damping of all AWs near the top of the core where the magnetic field should experience faster-than-exponential decay with radius. Thus, both axisymmetric and nonaxisymmetric dipole mixed modes should have suppressed amplitudes in stars with sufficiently strong core magnetic fields, and our simulations do not provide a mechanism explaining the g-mode character of partially-suppressed mixed modes described in \citet{Mosser2017}.

These results bolster recent efforts \citep{lecoanet_asteroseismic_2022,Rui2025,Duguid2024} to infer the strength of magnetic fields in astrophysical bodies based on wave suppression. We find that nonaxisymmetric IGWs convert to SM waves where the magnetic field strength reaches only a fraction of the value needed for the conversion of axisymmetric IGWs (we observe that for the same background field strength, the IGW$\rightarrow$SM turning points are higher in the nonaxisymmetric case; see Figure \ref{fig:evals}\textit{a},\textit{b}). Thus, we expect the critical field strength deduced from the axisymmetric problem \citep{lecoanet_conversion_2017} to provide a good lower bound on the magnetic field needed to account for the suppression of both axisymmetric and nonaxisymmetric dipole modes.

Future work should refine our results with numerical simulations in spherical geometry. The Cartesian model used here simplifies the problem and facilitates the derivation of the WKB amplitude equation, which is essential for comparing theory to simulation. Though the Cartesian wave modes discussed in this work largely resemble the spherical modes found by \cite{rui_gravity_2023}, some qualitative differences exist. For example, \cite{rui_gravity_2023} find (what we interpret here as) mixed SM-AW modes with large-scale oscillations confined to an equatorial band between Alfv\'en critical latitudes. It is unclear how much this equatorial focusing effect changes the wave interaction problem, since the behavior of the SM-AW modes in Cartesian and spherical geometries is otherwise similar. \rev{Simulations with different background magnetic field configurations may yield additional insight, particularly in the case for which the toroidal field component is much stronger than the poloidal components \citep[e.g.,][]{braithwaite_axisymmetric_2009}. Finally, f}urther investigations should also account for rotation, which acts differently on prograde versus retrograde modes (i.e., modes with $m$ of opposite sign, \citealt{lee_lowfrequency_1997}). Though rotational and geometric effects may introduce the differences described above, we expect our main conclusions to hold: energy carried inward by both axisymmetric and nonaxisymmetric IGWs is ultimately lost via interactions with a strong magnetic field. 

%% file: appendix.tex
\section{Dimensionless equations}\label{app:ndim}
The numerical simulations and WKB analysis use a dimensionless form of the magneto-Boussinesq equations (\ref{eqn:magbouss}) that are derived as follows. Quantities denoted with tildes ($\tilde{\;\;}$) are dimensionless. We scale time as $t = \omega^{-1} \tilde{t}$ and the background magnetic field as $\boldsymbol{B}_0 = \mathcal{B} \tilde{\boldsymbol{v}}_A =\mathcal{B}(\tilde{v}_{A x} \boldsymbol{e}_x + \tilde{v}_{A z} \boldsymbol{e}_z) $. Based on the dispersion relation for pure IGWs (\ref{eqn:igwdisprel}), we anticipate large horizontal length-scales (comparable to forcing wavelength $L$) and small vertical scales (comparable to $l_z = L \omega /N$). Accordingly, we scale distances anisotropically: $(x,y) = (L\tilde{x},L\tilde{y})$, $z = l_z \tilde{z}$, and $k_y = \tilde{k}_y/L$. The velocity and magnetic field perturbations are also scaled anisotropically to balance their respective divergence-free constraints (\ref{eqn:cont},\ref{eqn:gauss}). We decompose each as $\boldsymbol{u} = \boldsymbol{u}_H + w\boldsymbol{e}_z$ and  $\boldsymbol{b} = \boldsymbol{b}_H + b_z\boldsymbol{e}_z$, and let $\boldsymbol{u}_H = \mathcal{U} \tilde{\boldsymbol{u}}_H$, $w = \mathcal{U} (l_z/L) \tilde{w}$, $\boldsymbol{b}_H = \mathcal{U} \sqrt{\mu_0 \bar{\rho}} \tilde{\boldsymbol{b}}_H$, and $b_z = \mathcal{U} (l_z/L) \sqrt{\mu_0 \bar{\rho}} \tilde{b}_z$, where $ \mathcal{U}$ is the (velocity) amplitude of the IGWs. The mass equation (\ref{eqn:lindensity}) is balanced by scaling the density perturbation as $\rho = \bar{\rho} N^2 l_z \mathcal{U}/(g \omega L)\tilde{\rho}=(\bar{\rho} N {\mathcal{U}}/g) \tilde{\rho}$. Finally, the pressure perturbation is scaled to reflect leading-order hydrostatic balance in the vertical momentum equation: $p = \bar{\rho}N^2 l_z^2\mathcal{U}/(\omega L) \tilde{p}= (\bar \rho \omega\mathcal{U}L)\tilde{p}$. After nondimensionalizing and dropping tildes ($\tilde{\;\;}$), the linearized equations become
\begin{subequations}\label{eqn:ndimmagbouss}
    \begin{equation}
        \partial_t \boldsymbol{u}_H + \boldsymbol{\nabla}_H p -  \Gamma  v_{A z} \partial_z \boldsymbol{b}_H = -\textit{Fr}^2\Gamma  v_{A z}  \boldsymbol{\nabla}_H b_z +  \textit{Fr} \Gamma (\boldsymbol{\nabla}_H \times \boldsymbol{b}_H) \times v_{A x} \boldsymbol{e}_x - D\boldsymbol{u}_H,
    \end{equation}
    \begin{equation}
        \partial_z p +  \rho= -\textit{Fr}^2\partial_t w   + \textit{Fr} \Gamma v_{A x}(\textit{Fr}^2\partial_x b_z -   \partial_z b_x) - Dw,
    \end{equation}
    \begin{equation}
        \boldsymbol{\nabla}_H \boldsymbol{\cdot} \boldsymbol{u}_H +  \partial_z w = 0,
    \end{equation}
    \begin{equation}
        \partial_t \rho - w =  F -D\rho,
    \end{equation}
    \begin{multline}
        \partial_t \boldsymbol{b}_H - \Gamma v_{A z} \partial_z \boldsymbol{u}_H - \textit{Lu}_x^{-1} \textit{Fr}^2\partial_x^2\boldsymbol{b}_H -  \textit{Lu}^{-1}\left(\textit{Fr}^2\partial_y^2 +\partial_z^2\right)\boldsymbol{b}_H = -\textit{Fr}^2\Gamma  w \partial_\mathcal{Z}v_{A x}\boldsymbol{e}_x\\ +\textit{Fr} \Gamma \left(\boldsymbol{v}_A \boldsymbol{\cdot} \boldsymbol{\nabla}_H \boldsymbol{u}_H -\boldsymbol{u}_H \boldsymbol{\cdot} \boldsymbol{\nabla}_H v_{A x}\boldsymbol{e}_x\right) - D \boldsymbol{b}_H,
    \end{multline}
    \begin{equation}
        \partial_t b_z - \Gamma(v_{A z}\partial_z w - u\partial_x v_{A z}) - \textit{Lu}_x^{-1} \textit{Fr}^2\partial_x^2 b_z -  \textit{Lu}^{-1}\left(\textit{Fr}^2\partial_y^2 + \partial_z^2\right) b_z=  \textit{Fr} \Gamma  (v_{A x}\partial_x w - w \partial_\mathcal{Z}v_{A z}) - D b_z,
    \end{equation}
    \begin{equation}
        \boldsymbol{\nabla}_H \boldsymbol{\cdot} \boldsymbol{b}_H +  \partial_z b_z = 0,
    \end{equation}
\end{subequations}
where $\boldsymbol{\nabla}_H = \boldsymbol{e}_x \partial_x + \boldsymbol{e}_y \partial_y$, and where we have assumed that the background field is current-free (i.e., $\partial_{\mathcal{Z}} v_{A x} - \partial_x v_{A z} = 0$). For anisotropic diffusion (IVP III), $\textit{Lu}_x \neq \textit{Lu}$. For the remainder of the Appendix, \rev{we neglect the forcing and damping terms. That is, we set the functions $F$ and $D$ (defined in Equations \ref{eqn:forcing} and \ref{eqn:damping}, respectively) to zero}. All quantities are dimensionless and will be written without tildes ($\tilde{\;\;}$).

Finally, we observe that, in the absence of diffusion, the horizontally-integrated wave energy is governed by
\begin{multline}\label{eqn:horizavgenergy}
    {\partial_t}\int_{-\pi/k_y}^{-\pi/k_y} \int_0^1 \frac{1}{2}\left(\boldsymbol{u}_H \cdot \boldsymbol{u}_H + \textit{Fr}^2 w^2 + {\rho^2} + \boldsymbol{b}_H \cdot \boldsymbol{b}_H + \textit{Fr}^2 b_z^2\right) \mathrm{d}x\mathrm{d}y \\ +  \partial_z \int_{-\pi/k_y}^{-\pi/k_y} \int_0^1  \left[p w - \Gamma(\boldsymbol{b}_H \cdot \boldsymbol{u}_H) v_{A z} + \Gamma \textit{Fr} b_x v_{A x} w \right] \mathrm{d}x\mathrm{d}y= 0,
\end{multline}
which is used to derive an expression for the WKB group velocity in the following section.

\section{Development of the reduced equations}\label{app:wkb}
We set $\epsilon = \textit{Fr}$ and expand each field  $u$, $v$, $w$, $p$, $\rho$, $b_x$, $b_y$, $b_z$ as, e.g.,
\begin{equation}
    p = \Re\left\{\hat{p}(x, \mathcal{Z})\exp\left(i \int k_z(\mathcal{Z})\mathrm{d}z  +i k_y y - i t\right)\right\},
\end{equation}
with 
\begin{equation}\label{eqn:wkbexpansion}
    \hat{p}(x,\mathcal{Z}) = A(\mathcal{Z}) p_0(x, \mathcal{Z}) + \epsilon p_1(x, \mathcal{Z}) +  \epsilon^2 p_2(x, \mathcal{Z}) +...
\end{equation}
where the amplitude function $A$, vertical wavenumber $k_z$, and each $p_i$ vary with the ``slow'' vertical coordinate $\mathcal{Z} = \epsilon z$. Note that $k_z$ here is nondimensionalized by $l_z$. Then, the derivative operators transform as, e.g., 
$\partial_t p = -i \hat{p} \exp(i \Phi)$, $\partial_x p = \partial_x \hat{p} \exp(i \Phi)$, $\partial_y p = i k_y \hat{p} \exp(i \Phi)$, and finally, $\partial_z p = i k_z({\cal Z})  \hat{p} \exp(i \Phi) + \epsilon \partial_{\mathcal{Z}} \hat{p} \exp(i \Phi)$, where $\Phi = \int k_z(\mathcal{Z})\mathrm{d}z + k_y y - t$.

We substitute these ans\"atze into (\ref{eqn:ndimmagbouss}) and analyze the distinguished limit $\Gamma = O(1)$ as $\epsilon \to 0$. From the left-hand side of (\ref{eqn:ndimmagbouss}) we deduce the leading order resistive equations,
\begin{subequations}\label{eqn:order1}
    \begin{equation}
        -i \boldsymbol{u}_{0 H} + \hat{\boldsymbol{\nabla}} p_0 - i \Gamma v_{A z}k_z\boldsymbol{b}_{0 H} = \boldsymbol{0},
    \end{equation}
    \begin{equation}\label{eqn:zmomorder1}
        i k_z p_0 + \rho_0 = 0,
    \end{equation}
    \begin{equation}
        \hat{\boldsymbol{\nabla}} \boldsymbol{\cdot} \boldsymbol{u}_{0 H} + i k_z w_0 = 0,
    \end{equation}
    \begin{equation}\label{eqn:massorder1}
        -i \rho_0 - w_0= 0,
    \end{equation}
    \begin{equation}\label{eqn:inductionorder1}
        -i \boldsymbol{b}_{0 H} - i \Gamma k_z v_{A z} \boldsymbol{u}_{0 H} -\textit{Lu}_x^{-1}\textit{Fr}^2\partial_x^2 \boldsymbol{b}_{0 H} + \textit{Lu}^{-1}(\textit{Fr}^2 k_y^2+k_z^2)\boldsymbol{b}_{0 H}= \boldsymbol{0},
    \end{equation}
\end{subequations}
where $\hat{\boldsymbol{\nabla}} = \boldsymbol{e}_x \partial_x +  \boldsymbol{e}_y i k_y$, $\boldsymbol{u}_{0 H} = u_0 \boldsymbol{e}_x + v_0 \boldsymbol{e}_y$, and $\boldsymbol{b}_{0 H} = b_{0 x} \boldsymbol{e}_x + b_{0 y} \boldsymbol{e}_y$. Note that, while the diffusion terms in (\ref{eqn:order1}) are formally higher order in $\textit{Fr}$, we retain them in the equations for regularization.

In the absence of diffusion ($\textit{Lu}_x^{-1} = \textit{Lu}^{-1} = 0$), the leading order system (\ref{eqn:order1}) can be combined into a single homogeneous equation, $L_{k_z}p_0 = 0$ where 
\begin{equation}
    L_{k_z}(\cdot) = k_z^2 (\cdot) + \hat{\boldsymbol{\nabla}} \boldsymbol{\boldsymbol{\cdot}} \left( \frac{\hat{\boldsymbol{\nabla}} (\cdot)}{1-\Gamma^2 v_{A z}^2 k_z^2 } \right).
\end{equation}

The remaining piece of the leading-order WKB solution is the amplitude function $A(\mathcal{Z})$, which is obtained at the next order. The $O(\epsilon)$ equations form an inhomogeneous linear system that may be combined into the single equation
\begin{equation}\label{eqn:ordereps}
    L_{k_z}p_1 = f,
\end{equation}
where
\begin{subequations}
    \begin{equation}
        f = -i\partial_{\mathcal{Z}}(A w_ 0) +i {k_z}\partial_{\mathcal{Z}} (A p_ 0) - {k_z^2}\Gamma A v_{A x} b_{0 x} +  \hat{\boldsymbol{\nabla}}  \cdot \left( \frac{\boldsymbol{h}}{1-\Gamma^2 k_z^2 v_ {A z}^2}\right),
    \end{equation}
    \begin{equation}
        \boldsymbol{h} =\Gamma v_ {A z} \partial_{\mathcal{Z}}(A \boldsymbol{b}_{0 H} )+ \Gamma ( \hat{\boldsymbol{\nabla}} \times A \boldsymbol{b}_{0 H})\times v_ {A x}\boldsymbol{e}_x -  {\Gamma^2 v_ {A z} k_z} \left[v_ {A z} \partial_{\mathcal{Z}} (A \boldsymbol{u}_{0 H}) + v_ {A x} \partial_x(A \boldsymbol{u}_{0 H}) - A u_ 0 \partial_x v_ {A x} \boldsymbol{e}_x \right].
    \end{equation}
\end{subequations}
A solvability condition on the inhomogeneous term $f$ is required to ensure that a bounded solution exists, and may be obtained by taking the inner product of (\ref{eqn:ordereps}) with an element of the kernel of the adjoint operator $L_{k_z}^\dagger$. We observe that $L_{k_z}$ is self-adjoint up to complex conjugation of $k_z$ (i.e., $L_{k_z}^\dagger = L_{k_z^\ast}$) for the inner product $\langle a,b\rangle = \int_0^1 a^\ast b \mathrm{d}x$. Using this fact, we have that $p_0^\ast$ is in the kernel of $L_{k_z}^\dagger$; that is,   $L_{k_z}^\dagger p_0^\ast = L_{k_z^\ast}p_0^\ast =(L_{k_z}p_0)^\ast = 0$. Thus, taking the inner product of  (\ref{eqn:ordereps}) with $p_0^\ast$ yields the desired solvability condition:
\begin{equation}\label{eqn:solvcond}
    0 = \langle  p_0^\ast, f \rangle .
\end{equation}
Equation (\ref{eqn:solvcond}) simplifies to a first order differential equation in $\mathcal{Z}$ which determines the amplitude function up to a multiplicative constant:
\begin{equation}\label{eqn:amp}
    A(\mathcal{Z}) = \left\{\int_0^1 \left[p_0 w_0 + \Gamma v_{A z} (b_{0 x} u_0 - b_{0 y} v_0)\right]\mathrm{d}x \right\}^{-1/2}.
\end{equation}
A detailed derivation of (\ref{eqn:amp}) may be found in \rev{Appendix A.3.3 of }\cite{david_interaction_2024}.

Finally, if the amplitude $A$ is allowed to vary slowly with time $\mathcal{T} = \epsilon t$ (before the solution equilibrates), then one may show (by substituting the WKB ans\"atze into Equation \ref{eqn:horizavgenergy} and keeping only the leading order terms) that the WKB wave energy
\begin{equation}
    E_0 =\frac{\lvert A(\mathcal{Z},\mathcal{T}) \rvert^2}{4}\int_0^1 \left( {\boldsymbol{u}_{0 H}\cdot\boldsymbol{u}_{0 H}^\ast} + {\rho_0 \rho_0^\ast} + {\boldsymbol{b}_{0 H}\cdot\boldsymbol{b}_{0 H}^\ast}\right) \mathrm{d}x
\end{equation}
is transported vertically (in the absence of diffusion) according to
\begin{equation}
    (\partial_{\mathcal{T}} + c_{g z} \partial_\mathcal{Z}) E_0 = -E_0 \partial_\mathcal{Z} c_{g z},
\end{equation}
where the vertical WKB group velocity is given by
\begin{equation}\label{eqn:groupvel}
    c_{g z} = \frac{\text{(rate of) pressure work} + \text{Poynting flux}}{\text{wave energy}} = \frac{\int_0^1 \Re\left\{p_0^\ast {w}_ 0- \Gamma v_{A z} {\boldsymbol{b}}_ {0 H} \cdot {\
\boldsymbol{u}}_{0 H}^\ast\right\} \mathrm{d}x}{\frac{1}{2}\int_0^1 \left( {\boldsymbol{u}_{0 H}\cdot\boldsymbol{u}_{0 H}^\ast} + {\rho_0 \rho_0^\ast} + {\boldsymbol{b}_{0 H}\cdot\boldsymbol{b}_{0 H}^\ast}\right) \mathrm{d}x}.
\end{equation}

\section{Eigenvalue problem implementation in Dedalus}\label{app:evp}
\subsection{Ideal axisymmetric eigenproblem}\label{app:evp:idealaxi}
The WKB wave modes and vertical wavenumbers are found numerically using the Dedalus pseudospectral code. For the ideal ``axisymmetric'' case  ($\textit{Lu}_x^{-1} = \textit{Lu}^{-1} = 0$, $k_y =0$) in Figure \ref{fig:evals}\textit{a}, we solve the following generalized eigenvalue problem
\begin{equation}\label{eqn:idealevp}
    \begin{bmatrix}
        0 & i\partial_x & - k_y \\
        i \partial_x & 1 & 0 &  \\
       -  k_y & 0 & 1 & 
    \end{bmatrix}
    \begin{bmatrix}
        p_0 \\ u_0 \\ v_0
    \end{bmatrix}
    +k_z^2
    \begin{bmatrix}
        1 & 0 & 0 \\
        0&-\Gamma^2  v_{A z}^2 & 0 \\
        0&0 & -\Gamma^2  v_{A z}^2
    \end{bmatrix}
    \begin{bmatrix}
        p_0 \\ u_0 \\ v_0
    \end{bmatrix}
    = \boldsymbol{0}
\end{equation}
for eigenvalue $k_z^2$ and eigenvector $[p_0 \;u_0 \; v_0 ]^T$. The system (\ref{eqn:idealevp}) only depends on $\mathcal{Z}$ parametrically (through $v_{A z}$) and may be solved over one dimension ($x$) at each level in $\mathcal{Z}$. We use Dedalus’ dense solver to find the eigenpairs using $N_x = 96$ complex Fourier modes in $x$ at each of $N_z = 512$ uniformly spaced values of $\mathcal{Z} \in [0,0.25]$. To assess numerical convergence, the process is repeated with $N_x = 128$ complex Fourier modes, and only the eigenvalues that differ by less than $10^{-8}$ times their original value are kept. 

\subsection{Resistive nonaxisymmetric eigenproblem}\label{app:evp:resnonaxi}
A different procedure yields the resistive wave modes used to construct the nonaxisymmetric ($k_y=2\pi$) WKB solution in Section \ref{sec:wkb}. We find the generalized eigenvalues $k_z$ (see Figure \ref{fig:evals}\textit{b},\textit{c}) and eigenvectors $\boldsymbol{q}_0 = [u_0 \;v_0\; w_0\;p_0\;\rho_0\;b_{0 x}\;b_{0 y}\;b_{0 x,z}\;b_{0 y,z}]^T$ of
\begin{equation}\label{eqn:resistiveevp}
     \begin{bmatrix}
        1 & 0 & 0 & i\partial_x & 0 & 0 & 0 & 0\\
        0 & 1 & 0 & -k_y & 0 & 0 & 0 & 0\\
        0 & 0 & -1 & 0 & 0 & 0 & 0 & 0\\
        i\partial_x & -k_y & 0 & 0 & 0 & 0 & 0 & 0\\
        0 & 0 & 0 & 0 &1 -i\hat{D} & 0 & 0 & 0\\
        0 & 0 & 0 & 0 & 0 & 1 -i\hat{D} & 0 & 0\\
        0 & 0 & 0 & 0 & 0 & 0 & 1 & 0\\
        0 & 0 & 0 & 0 & 0 & 0 & 0 & 1\\
    \end{bmatrix}
    \begin{bmatrix}
        u_0 \\v_0\\ w_0\\p_0\\b_{0 x}\\b_{0 y}\\b_{0 x,z}\\b_{0 y,z}
    \end{bmatrix} + k_z
    \begin{bmatrix}
        0 & 0 & 0 & 0 & \Gamma v_{A z} & 0 & 0 & 0\\
        0 & 0 & 0 & 0 & 0 & \Gamma v_{A z} & 0 & 0\\
        0 & 0 & 0 & -1 & 0 & 0 & 0 & 0\\
        0 & 0 & -1 & 0 & 0 & 0 & 0 & 0\\
        \Gamma v_{A z} & 0 & 0 & 0 & 0 & 0 & \textit{Lu}^{-1} & 0\\
        0 & \Gamma v_{A z} & 0 & 0 & 0 & 0 & 0 & \textit{Lu}^{-1}\\
        0 & 0 & 0 & 0 & -i & 0 & 0 & 0\\
        0 & 0 & 0 & 0 & 0 & -i & 0 & 0\\
    \end{bmatrix}
    \begin{bmatrix}
        u_0 \\v_0\\ w_0\\p_0\\b_{0 x}\\b_{0 y}\\b_{0 x,z}\\b_{0 y,z}
    \end{bmatrix}
    = \boldsymbol{0}   
\end{equation}
where $\hat{D} = \textit{Lu}_x^{-1}\textit{Fr}^2\partial_x^2 - \textit{Lu}^{-1}\textit{Fr}^2k_y^2$ is the transformed horizontal diffusion operator. The system (\ref{eqn:resistiveevp}) is formed by eliminating $\rho_0$ from (\ref{eqn:order1}) and appending the equations $b_{0 x,z} = i k_z b_{0 x}$ and $b_{0 y,z} = i k_z b_{0 y}$.

In the presence of diffusion,  (\ref{eqn:resistiveevp}) yields four branches of ``dipolar'' wavenumbers $k_z$ (i.e., those for which $\lvert p_0\rvert$ has two zeros on $x \in [0,1)$) that vary continuously over $\mathcal{Z} \in [0,0.25]$. Our procedure for finding the eigenpairs along these branches is as follows. First, we set $\mathcal{Z}=0$ in  (\ref{eqn:resistiveevp}) and apply Dedalus' dense solver at two different resolutions ($N_x = 96$ and $N_x=128$ complex Fourier modes in $x$). At this height, all four eigenvalues correspond to evanescent waves. Next, the resolved eigenvalue for each of the four dipolar branches is used as the target for Dedalus’ sparse solver with $N_x = 1024$ complex Fourier modes. Then, we increment $\mathcal{Z}$ by $\Delta  \mathcal{Z} = 0.25/N_z$ (with $N_z = 3072$) and use the eigenvalue from the previous level in $\mathcal{Z}$ as the target for the sparse solver at the current level. The process is repeated up to $\mathcal{Z} = 0.25$, keeping only the mode with the same parity as the previous level. Thus, at every $j$th level, we keep one eigenpair with eigenvalue $k_{z,j}$ and eigenfunctions $\boldsymbol{q}_{0,j} = [u_{0,j}(x) \;\;v_{0,j}(x)\;\; w_{0,j}(x)\;\;p_{0,j}(x) \dots]^T$. At this stage, the set of eigenfunctions $\boldsymbol{q}_{0,j}$ for each of the four branches is normalized such that $p_{0,j}(x=1/3)$ is constant for all $j=1,2,\dots, N_z$ (this ensures continuity over $\mathcal{Z}$).

The four wavenumber branches have turning points $\mathcal{Z}_t$ (which we identify as maxima in $\lvert \mathrm{d} k_z/ \mathrm{d} \mathcal{Z}\rvert$), where the assumption that $\mathrm{d} k_z/ \mathrm{d} \mathcal{Z} = O(\epsilon)$ is violated. Thus, the WKB solution is only asymptotically exact away from the turning points, and these heights naturally delimit different WKB wave modes (e.g., a turning point separates the sine-parity IGW mode, IGW-0, from the associated evanescent mode, evan.-0). In addition, the cosine-parity SM mode (SM-1) is separated from the mixed SM-AW mode (SM-AW-1) at the height $\mathcal{Z}_{A B}$ at which their vertical wavenumbers intersect the Alfv\'en wavenumber boundary, $k_z = k_{A B}$ (see Figure \ref{fig:evals}\textit{b}, where solid purple and dashed purple curves intersect the edge of the gray shaded region).

The turning points $\mathcal{Z}_t$ and Alfv\'en boundary intersection height $\mathcal{Z}_{A B}$ split the four continuous branches into 11 separate wave modes. Two of these modes have $\Im\{k_z\} > 0$ at $\mathcal{Z} =0$ such that the waves grow exponentially with depth near the bottom of the domain. We discard these two unphysical evanescent modes, leaving the nine wave modes that comprise the WKB solution in Section \ref{sec:wkb}. The first branch is split into evan.-0 (for $0 \le \mathcal{Z} < \mathcal{Z}_{t,\text{IGW-0}} \approx 0.0522$) and IGW-0 (for $\mathcal{Z} \ge \mathcal{Z}_{t,\text{IGW-0}}$); the second branch is split into evan.-1 (for $0 \le \mathcal{Z} < \mathcal{Z}_{t,\text{IGW-1}} \approx 0.0461$) and IGW-1 (for $\mathcal{Z} \ge \mathcal{Z}_{t,\text{IGW-1}}$); the third branch is split into SM-0 (for $\mathcal{Z}_{t,\text{IGW-0}} \le \mathcal{Z} < \mathcal{Z}_{t,\text{AW-0}} \approx 0.163$) and AW-0 (for $\mathcal{Z} \ge \mathcal{Z}_{t,\text{AW-0}}$); and the fourth branch is split into SM-1 (for $\mathcal{Z}_{t,\text{IGW-1}} \le \mathcal{Z} < \mathcal{Z}_{A B} \approx 0.0496$), SM-AW-1 (for $\mathcal{Z}_{A B} \le \mathcal{Z} < \mathcal{Z}_{t,\text{AW-1}} \approx 0.0910$), and AW-1 (for $\mathcal{Z} \ge \mathcal{Z}_{t,\text{AW-1}}$).

For each of the nine waves, the full eigenmodes $\boldsymbol{q}_0(x,\mathcal{Z}) = [u_0(x,\mathcal{Z}) \;\;v_0(x,\mathcal{Z})\;\; w_0(x,\mathcal{Z})\;\;p_0(x,\mathcal{Z}) \dots]^T$ are formed by ``stitching'' together the eigenfunctions $\boldsymbol{q}_{0,j}(x) = [u_{0,j}(x) \;\;v_{0,j}(x)\;\; w_{0,j}(x)\;\;p_{0,j}(x) \dots]^T$ at each $\mathcal{Z}_j$ over the vertical extent of the wave. The eigenmodes for IGW-0 and IGW-1 are normalized such that $\langle p_0,p_0\rangle=1$ at the forcing height $\mathcal{Z}_0$ with $p_{0,\text{IGW-0}} = 1$ at $(x,\mathcal{Z})=(0,\mathcal{Z}_0)$ and $p_{0,\text{IGW-1}} = 1$ at $(x,\mathcal{Z})=(1/4,\mathcal{Z}_0)$. This ensures that $p_{0,\text{IGW-0}} + i p_{0,\text{IGW-1}} \approx \exp(i k_x x)$ at $\mathcal{Z} = \mathcal{Z}_0$ to match the forcing. Then, the eigenmodes for SM-0 and SM-1 are rescaled such that $p_{0,\text{SM-0}} \approx p_{0,\text{IGW-0}}$ at $(x,\mathcal{Z}) = (0,\mathcal{Z}_{t,\text{IGW-0}})$ and $p_{0,\text{SM-1}} \approx p_{0,\text{IGW-1}}$ at $(x,\mathcal{Z)} = (1/4,\mathcal{Z}_{t,\text{IGW-1}})$. (Since we assume perfect conversion from IGWs to SM waves, the eigenmodes must match across the turning points). Finally, the eigenmodes for the remaining waves are rescaled such that for every pair of adjacent wave modes from the same branch (e.g., evan.-0 and IGW-0, but not evan.-0 and SM-0), $p_0$ is continuous across the associated transition height (either a turning point or $\mathcal{Z}_{A B}$).

\section{WKB solution}\label{app:wkbsoln}
The amplitude function $A(\mathcal{Z})$ for each of the nine resistive wave modes is calculated using the ideal amplitude equation (\ref{eqn:amp}), and the WKB solution is computed by summing the modes as, e.g.,
\begin{equation}\label{eqn:uwkb}
    u_{\text{WKB}} = \Re\left\{C\sum_{n=1}^9  \mathcal{A}_n(\mathcal{Z})u_{0,n}(x, \mathcal{Z})\exp\left(i\theta_n(z) +i k_y y - i t\right)\right\},
\end{equation}
with
\begin{subequations}
    \begin{equation}
        \mathcal{A}_n(\mathcal{Z}) = c_n A_n(\mathcal{Z}),
    \end{equation}
    \begin{equation}\label{eqn:phaseintergral}
        \theta_n(z) = \int_{z_{b,n}}^{z}k_{z,n}(\mathcal{Z})\mathrm{d}z' + \Delta\theta_n,
    \end{equation}
\end{subequations}
where the lower limit of integration $z_{b}$ is set to the lower bound of each wave mode (e.g., $z_{b} = \mathcal{Z}_{t,\text{IGW-0}}/\textit{Fr}$ for IGW-0). Importantly, the amplitude equation (\ref{eqn:amp}) and eigenvalue problem (\ref{eqn:resistiveevp}) only define $\mathcal{A}_n$ and $\theta_n$ up to multiplicative and additive constants, respectively. Thus, relative amplitudes $c_n$ (which are distinct from the overall complex amplitude $C$) and the phase shifts $\Delta\theta_n$ must be determined independently.

For IGW-0 and IGW-1, $c_n$ and $\Delta\theta_n$ are set such that the IGW modes match the forcing in the simulation at $\mathcal{Z}_0 = 0.225$:
\begin{subequations}
    \begin{equation}
        \theta_{\text{IGW-0}}(\mathcal{Z}_0/\textit{Fr}) = \theta_{\text{IGW-1}}(\mathcal{Z}_0/\textit{Fr}) -\pi/2 =0,
    \end{equation}
    \begin{equation}
        \mathcal{A}_{\text{IGW-0}}(\mathcal{Z}_0) = \mathcal{A}_{\text{IGW-1}}(\mathcal{Z}_0) = 1.
    \end{equation}
\end{subequations}
We assume total conversion from IGWs to SM waves, and thus set each $c_n$ for SM-0, SM-1, evan.-0, and evan.-1 such that
\begin{subequations}
    \begin{equation}
        \mathcal{A}_{\text{SM-0}}(\mathcal{Z}_{t,\text{IGW-0}})= \mathcal{A}_{\text{evan.-0}}(\mathcal{Z}_{t,\text{IGW-0}}) = \mathcal{A}_{\text{IGW-0}}(\mathcal{Z}_{t,\text{IGW-0}}),
    \end{equation}
    \begin{equation}
        \mathcal{A}_{\text{SM-1}}(\mathcal{Z}_{t,\text{IGW-1}})= \mathcal{A}_{\text{evan.-1}}(\mathcal{Z}_{t,\text{IGW-1}}) = \mathcal{A}_{\text{IGW-1}}(\mathcal{Z}_{t,\text{IGW-1}}).
    \end{equation}
\end{subequations}
At the IGW$\rightarrow$SM turning points, the phase differences are assumed to match the theoretical values found in the ``axisymmetric'' case by \cite{lecoanet_conversion_2017} such that
\begin{subequations}
    \begin{equation}
        \theta_{\text{SM-0}}(\mathcal{Z}_{t,\text{IGW-0}}/\textit{Fr}) +\pi/2 =\theta_{\text{evan.-0}}(\mathcal{Z}_{t,\text{IGW-0}}/\textit{Fr})=  \theta_{\text{IGW-0}}(\mathcal{Z}_{t,\text{IGW-0}}/\textit{Fr}),
    \end{equation}
    \begin{equation}
        \theta_{\text{SM-1}}(\mathcal{Z}_{t,\text{IGW-1}}/\textit{Fr}) +\pi/2 =\theta_{\text{evan.-1}}(\mathcal{Z}_{t,\text{IGW-1}}/\textit{Fr})=  \theta_{\text{IGW-1}}(\mathcal{Z}_{t,\text{IGW-1}}/\textit{Fr})
    \end{equation}
\end{subequations}
Further, we assume total conversion from SM-0 to AW-0, SM-1 to SM-AW-1, and SM-AW-1 to AW-1 with no associated phase shifts:
\begin{subequations}
    \begin{equation}
        \mathcal{A}_{\text{AW-0}}(\mathcal{Z}_{t,\text{AW-0}})= \mathcal{A}_{\text{SM-0}}(\mathcal{Z}_{t,\text{AW-0}}),\quad\theta_{\text{AW-0}}(\mathcal{Z}_{t,\text{AW-0}}/\textit{Fr})= \theta_{\text{SM-0}}(\mathcal{Z}_{t,\text{AW-0}}/\textit{Fr}),
    \end{equation}
    \begin{equation}
        \mathcal{A}_{\text{SM-AW-1}}(\mathcal{Z}_{AB})= \mathcal{A}_{\text{SM-1}}(\mathcal{Z}_{AB}),\quad \theta_{\text{SM-AW-1}}(\mathcal{Z}_{AB}/\textit{Fr})= \theta_{\text{SM-1}}(\mathcal{Z}_{AB}/\textit{Fr}),
    \end{equation}
    \begin{equation}
        \mathcal{A}_{\text{AW-1}}(\mathcal{Z}_{t,\text{AW-1}})= \mathcal{A}_{\text{SM-AW-1}}(\mathcal{Z}_{t,\text{AW-1}}),\quad \theta_{\text{AW-1}}(\mathcal{Z}_{t,\text{AW-1}}/\textit{Fr})= \theta_{\text{SM-AW-1}}(\mathcal{Z}_{t,\text{AW-1}}/\textit{Fr}).
    \end{equation}
\end{subequations}
Finally, the complex amplitude $C$ (which encodes the overall amplitude and phase of $u_{\text{WKB}}$) is determined by fitting the WKB solution to $u_{\text{IVP}}$ at $(x,\mathcal{Z}) = (1/3,0.175)$. To construct $v_{\text{WKB}}$, we simply replace each $u_{0,n}$ in (\ref{eqn:uwkb}) with the corresponding azimuthal velocity eigenfunction $v_{0,n}$.